\documentclass[letterpaper,superscriptaddress,aps,pre,nolongbibliography,twocolumn,showpacs,floatfix]{revtex4-2} 
\usepackage{graphicx,helvet}
\usepackage{color}
\usepackage{bm}                     
\usepackage{float}
\usepackage{tikz}
\usepackage{braket}
\usepackage{soul}
\usepackage{amsfonts}
\usepackage{amsmath}
\usepackage{lipsum}
\usepackage[percent]{overpic} 
\usepackage{MnSymbol}%

\usepackage{hyperref}
\usepackage{subfiles} 

\def\II{\hbox{$1\hskip -1.2pt\vrule depth 0pt height 1.6ex width 0.7pt\vrule depth 0pt height 0.3pt width 0.12em$}}

\newcommand{\zurich}{Department of Biosystems Science and Engineering, ETH Z\"urich, 4058 Basel, Switzerland}
\newcommand{\regens}{Institut f\"ur Theoretische Physik, Universit\"at Regensburg, 93040 Regensburg, Germany}
\newcommand{\lanzhou}{School of Physical Science and Technology, Lanzhou University, Lanzhou, 730000 Gansu, China}

\begin{document}
	\title{Universal S-matrix correlations for complex scattering of many-body wavepackets:\\ theory, simulation and experiment}
	\author{Andreas Bereczuk} \affiliation{\regens}
	\author{Barbara Dietz}
	\altaffiliation{email:  Dietz@lzu.edu.cn}
	 \affiliation{\lanzhou}
	\author{Jiongning Che} \affiliation{\lanzhou}
	\author{Jack Kuipers} \affiliation{\zurich}
	\author{Juan-Diego Urbina}
	\altaffiliation{email: Juan-Diego.Urbina@ur.de}
	 \affiliation{\regens}
	\author{Klaus Richter} \affiliation{\regens}

\begin{abstract} \small
We present an in-depth study of the universal correlations of scattering-matrix entries required in the framework of non-stationary many-body scattering where the incoming states are localized wavepackets. Contrary to the stationary case the emergence of universal signatures of chaotic dynamics in dynamical observables manifests itself in the emergence of universal correlations of the scattering matrix at different energies. We use a semiclassical theory based on interfering paths, numerical wave function based simulations and numerical averaging over random-matrix ensembles to calculate such correlations and compare with experimental measurements in microwave graphs, finding excellent agreement. Our calculations show that the universality of the correlators survives the extreme limit of few open channels relevant for electron quantum optics, albeit at the price of dealing with large-cancellation effects requiring the computation of a large class of semiclassical diagrams.
\end{abstract}

\maketitle

\section{Introduction}
Two major achievements in the field of quantum transport are the realization that non-interacting stationary transport can be described in terms of single-particle scattering, the Landauer-Büttiker approach \cite{datta_1995, imry2002introduction}, and that scattering through potentials supporting classical chaotic dynamics imprints universal signatures to quantum mechanical observables alluded to in the Bohigas-Giannoni-Schmidt conjecture \cite{Berry1979,Casati1980,Bohigas1984,Beenakker1997}. The combination of these two ideas allows for an extremely powerful description of interference phenomena in the conductance properties of chaotic and weakly disordered quantum systems that focus on the emergence of universal, robust and observable manifestations of quantum coherence \cite{Richter1999, casati2000new}. The key object of study is then the single-particle, energy-dependent scattering (S) - matrix giving the amplitudes of the process where a particle with energy $E$ is injected through the incoming channels and scattered into the outgoing ones. It is the dependence of the S-matrix on the energy of the incoming flux $S(E)$ and the statistical correlations among its entries, that in turn are directly connected with observables like conductance, shot noise, etc. \cite{Beenakker1997}.

During the last decades, two complementary and rigorously equivalent analytical approaches to describe these universal effects have emerged. On one side, one has the machinery of Random Matrix Theory (RMT) in its two possible formulations, the so-called Heidelberg approach where the internal dynamics of the chaotic scattering potential is modelled independently of the random couplings to the leads \cite{Mahaux1966}, and the Circular Ensembles approach where one directly constructs random S matrices constrained to respect the fundamental properties of unitarity and the microscopic symmetries of the system \cite{mehta2004random, guhr1998random}. On the other side, we have the semiclassical approximation based on sums over interfering amplitudes associated to classical trajectories joining the incoming and outgoing modes \cite{Richter1999,jalabert1990, waltner2012semiclassical,Richter2002}.

During the last years, as the interest in many-body interference exploded, driven by the realization that non-interacting scattering of bosonic states with large number of particles is a promising candidate to show quantum supremacy \cite{tillmann2013experimental, wang2017high, aaronson2011computational}, the RMT/semiclassical approach to single-particle complex scattering entered the arena of many-body scattering \cite{Urbina_2016, oliveira2020immanants}. The relevant physical observable in this context is the distribution of scattered particles as measured in the outgoing channels when the incoming state is a localized wavepacket \cite{tillmann2013experimental, wang2017high}. In such scenario we depart from the stationary scattering at fixed energy and the microscopic input of the theory requires connecting the S-matrix at different energies.

In this article we present an extensive study of the S-matrix cross-correlations at different energies required within the study of universal features on complex scattering of many-body wavepackets. Our study covers all aspects of the problem, by presenting the RMT description using large ensembles of random matrices, the semiclassical approach by summing over a very large number of diagrams made of interfering classical paths, simulations based on a numerically exact tight-binding approach, all showing excellent agreement with experimental measurements in microwave graphs. 

A major motivation of our work is the recent interest in the universal properties of many-body scattering in systems of non-interacting, identical particles, and specially their mesoscopic effects within the arena of electron quantum optics. The combination of these two physical regimes comes with two new features. First, the focus on a particular set of S-matrix correlations with an energy dependence and a combination of entry indices that has not been addressed before, while it is very characteristic of the many-body context. Second, the need to push the asymptotic limit of large number $N \gg 1$ of open channels, so typical of both RMT and semiclassical treatments, into the regime of $N = {\cal O}(1)$. We observe that, in fact, the $1/N$ expansion efficiently captures the dependence of the correlators on the control parameters in most cases, giving full confidence to the delicate semiclassical treatment. Remarkably, we also find that the cases where the semiclassical expansion depends on large cancellation effects that drastically slow down the speed of convergence, actually are also the cases where the signal/noise ratio of the experimental signal is the worst within the whole set of measurements.

The paper is organized as follows: First we link the four-point S-matrix correlations of interest to the non-stationary many-body wave packet scattering in Section \ref{sec:S-matrix correlations for the scattering of many-body wavepackets}. In Section \ref{sec:The semiclassical approach} the semiclassical approach for the S-matrix is introduced (Subsection \ref{subsec:semiclassical approach of quantum transport}) and followed by a possible application of that approach: an expansion of a two-point S-matrix correlator to first order in $1/N$ called the diagonal approximation. The principle proceeding to evaluate the energy dependent four-point correlators semiclassically is explained in Subsection \ref{subsec:Semiclassical calculation of energy dependent correlators}. In Section \ref{sec:Experiment} we link quantum chaos to quantum graphs and the experimentally realized microwave setup (\ref{subsec:Quantum graphs as model systems for quantum chaos}), which is described in \ref{subsec:Setup for energy dependent correlations}. The two numerical approaches in use, the tight-binding method and the Heidelberg approach, are depicted in Section \ref{sec:Numerical approaches}. The resulting energy dependent four-point correlator data of the experiment, numerical approaches and semiclassical analysis are condensed in Section \ref{sec:Results and discussion}. Finally, the conclusions are drawn in \ref{sec:Conclusions}.


\section{S-matrix correlations for the scattering of many-body wavepackets}
\label{sec:S-matrix correlations for the scattering of many-body wavepackets}
\begin{figure}[t]
\centering
\includegraphics[width=0.6\columnwidth]{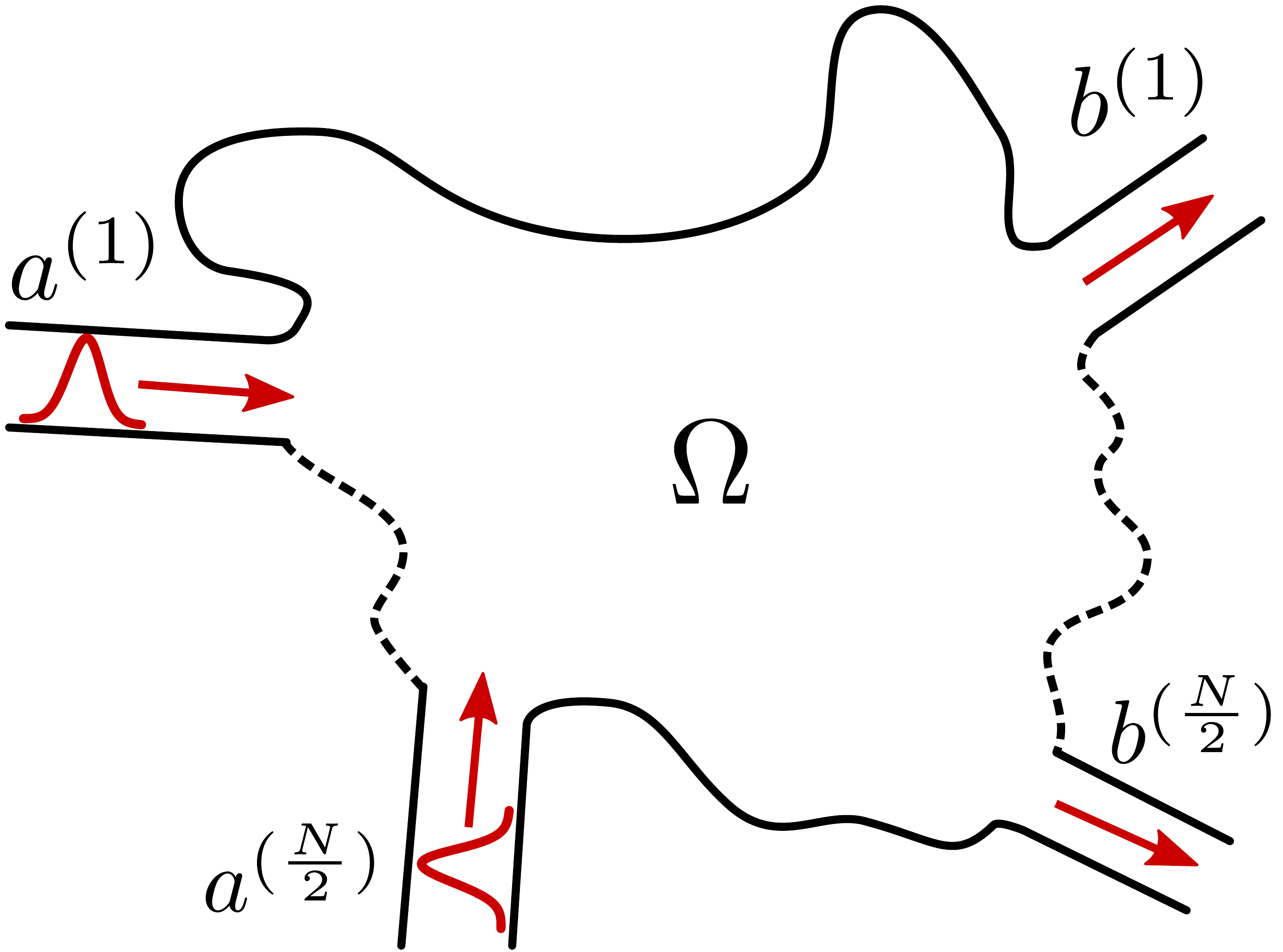}
\caption{Many-body scattering of indistinguishable particles by a cavity exhibiting single-particle chaos. The incoming many-body states are build by occupying localized wavepackets.} 
\label{fig:SP1}
\end{figure}

Within the standard approach to mesoscopic many-body scattering, described in Fig.~\ref{fig:SP1}, incoming particles ($i\!=\!1,\ldots,n$) with positions $(x_{i},y_{i})$ occupy single-particle states represented by normalized wavepackets 
\begin{equation}
\phi_{i}(x_{i},y_{i})\propto {\rm e}^{-\mathrm{i} k_{i} x_{i}}X_{i}(x_{i}-z_{i})\chi_{a_{i}}(y_{i}) \, 
\end{equation}
marking a difference with the stationary picture where one considers the limit of incoming states that are fully delocalized in position and approach asymptotically eigenstates of the momentum.

Along the longitudinal direction, the single-particle wavepacket $X_{i}(x_{i}-z_{i})$ occupied by the $i$th particle is described by three parameters. First, its variance $s^{2}_{i}$, second its mean initial position $z_{i} \gg s_{i}$, and third the mean momentum $\hbar k_{i}=mv_{i}>0$ with which it approaches the cavity. For the corresponding transverse wavefunction  in the incoming channel $a_{i}\!\in\!\{a^{(1)},\ldots,a^{(N/2)}\}$ we have the eigenstate of the transversal confinement, denoted by $\chi_{a_{i}}(y_{i})$ with energy $E_{a_{i}}$. 

For simplicity we assume that, except for their relative positions parametrized by $z_{ij}\!=\!z_{i}-z_{j}$, the wavepackets and initial transversal modes are identical: $s_{i}\!=\!s, {\rm \ }\hbar k_{i}\!=\!\hbar k\!=\!mv, {\rm \ and \ } E_{a_{i}}\!=\!E_{\rm ch} {\rm \ \ for \ all \ }i$. It is important to remark that, however, a more general treatment is possible.

The key observable we are interested in is the joint probability to find the $n$ particles, entering the cavity through channels ${\bf b}\!=\!(b_{1},\ldots,b_{n})$ with energies ${\bf E}\!=\!(E_{1},\ldots,E_{n})$, in channels ${\bf a}\!=\!(a_{1},\ldots,a_{n})$. Following the standard theory, this probability is given in terms of the ${\bf E}$-dependent $n$-particle amplitude  \cite{weinberg1995quantum,newton2013scattering}
\begin{equation}
\label{eq:Ampl}
A_{{\bf a},{\bf b}}({\bf E})\!=\!\prod_{i=1}^{n}\frac{{\rm e}^{-\mathrm{i}(k\!-\!k_{{\rm ch}}(E_{i}))z_{i}}}{\sqrt{\hbar v_{{\rm ch}}(E_{i})}}\tilde{X}(k-k_{{\rm ch}}(E_{i}))S_{b_{i},a_{i}}(E_{i}) \, 
\end{equation}
as
\begin{equation}
P_{{\bf a},{\bf b}}(E_{{\rm ch}})=|A_{{\bf a},{\bf b}}({\bf E})|^{2},
\end{equation}
where Eq.~(\ref{eq:Ampl}) formally defines the single-particle S-matrix $S_{b,a}(E)$ connecting the incoming and outgoing channels $a$ and $b$ at energy $E$. We also have $\hbar k_{{\rm ch}}(E)= mv_{{\rm ch}}(E)=\sqrt{2m(E-E_{{\rm ch}})}$ and $\tilde{X}(k)=(1/\sqrt{2\pi})\int_{-\infty}^{\infty}{\rm e}^{-\mathrm{i} kx}X(x)\mathrm{d}x$. 

If the particles are identical, quantum indistinguishability demands their joint state to be (anti-) symmetrized \cite{sakurai1995}. Denoting with $\epsilon=-1~(+1)$ for fermions (bosons), a symmetrized amplitude is obtained by summing over the $n!$ elements ${\cal P}$ of the permutation group,
\begin{equation}
\label{eq:Aeps}
A^{(\epsilon)}_{{\bf a},{\bf b}}({\bf E})=\sum_{{\cal P}}\epsilon^{{\cal P}}A_{{\bf a},{\cal P}{\bf b}}({\cal P}{\bf E}) \,  
\end{equation}
to get a many-body transition probability that is naturally separated into an incoherent contribution
\begin{equation}
\label{eq:Pinc}     
P^{{\rm inc}}_{{\bf a},{\bf b}}({\bf E})=\sum_{{\cal P}}|A_{{\bf a},{\cal P}{\bf b}}({\cal P}{\bf E})|^{2} \, ,
\end{equation}
and a many-body term sensitive to interference between different distinguishable configurations 
\begin{equation} 
\label{eq:Pint}    
P^{{\rm int}}_{{\bf a},{\bf b}}({\bf E}) =2n!\Re\sum_{{\cal P} \ne {\rm id.}}\epsilon^{{\cal P}}A_{{\bf a},{\cal P}{\bf b}}({\cal P}{\bf E}) A^{*}_{{\bf a},{\bf b}}({\bf E}) \, .
\end{equation}

We see that, while the transition probability for distinguishable particles, $P^{{\rm inc}}_{{\bf a},{\bf b}}$, is insensitive to the relative positions of the incoming wavepackets $z_{ij}$, for the indistinguishable situation it depends on the offsets $z_{ij}$ through its coherent contribution. This allows for an effective tuning of interference through dephasing, and thereby can be heuristically understood as sensitive to the degree of indistinguishability itself  \cite{tichy2014stringent,Ra1227,Broome794,Urbina_2016}. This type of effect is amplified by a further integration over the energies to obtain the transition probabilities in channel space. In bosonic systems, the dependence of this reduced probability on the offsets is the celebrated Hong-Ou-Mandel profile \cite{Hong1987} and its generalizations for $n>2$, while for fermions it is responsible for the Pauli dip, both experimentally accessible signatures of quantum indistinguishability.

The emergence of universal features in the complex scattering of many-body wavepackets is made explicit by introducing small energy windows to smooth the highly-oscillatory energy dependence of the transition probabilities. This smoothing affects mainly the energy dependence of the functions with strongest oscillatory dependence on the energies, that in chaotic systems is carried by the S-matrix. We conclude, therefore, that averages of many-body transition probabilities have universal signatures originating from that of the energy correlations of S matrices with certain combinations of energy differences and channels, as required by Eqs.~(\ref{eq:Ampl}),(\ref{eq:Aeps})-(\ref{eq:Pint}). 

Let us consider as an example the case $n=2$, corresponding to the standard Hong-Ou-Mandel (HOM) type of setup routinely used to verify the many-body coherence in quantum-optics scenarios. However, the setting considered here, cavity scattering, is much more complex than the use of a semi-transparent mirror in the original HOM proposal. Here, we have ${\bf E}=(E_{1},E_{2})$ and the group of permutations has only two elements ${\rm id},{\cal P}$. Making explicit the products in Eqs~(\ref{eq:Pinc},\ref{eq:Pint}), we end up with expressions involving the combinations  
\begin{equation}
\vert S_{b,a} (E_1) \vert^2 \; \vert S_{d,c} (E_2) \vert^2
\end{equation}
for the incoherent contribution, and 
\begin{eqnarray}
&&S_{b,a} (E_1) S_{d,c} (E_2) S^*_{d,a}   (E_2) S^*_{b,c} (E_1) 
\end{eqnarray}
for the interference one. Under energy smoothing, to be defined in \ref{subsec:Semiclassical calculation of energy dependent correlators}, these oscillating products are transformed into smooth correlation functions, that will carry their universal behavior for chaotic cavities over to the many-body transition probabilities. For reasons of completeness we will also study 
\begin{eqnarray}
&&S_{b,a} (E_1) S_{d,c} (E_1) S^*_{d,a}   (E_2) S^*_{b,c} (E_2) ,
\end{eqnarray}
which has been addressed before in \cite{Davis1989,dietz2010cross} in the context of cross-correlation functions.
The study of such correlations is the subject of this paper.



\section{The semiclassical approach}
\label{sec:The semiclassical approach}
\subsection{Semiclassical approach of quantum transport}
\label{subsec:semiclassical approach of quantum transport}
In the Landauer-B\"uttiker formalism \cite{landauer1970,buttiker1986,datta_1995} the S-matrix elements ${S_{b,a} (E)}$ are the probability amplitudes for a particle to elastically scatter from the incoming mode $a$ into the outgoing mode $b$, both at energy $E$. The size of the ${N \times N}$ matrix $S_{b,a}$ is given by the sum over open modes $N_i$ in the $m$ channels: ${N=\sum_{i=1}^m N_i}$. In the semiclassical limit, at large $N$ or when the system size is large compared to the Fermi wavelength, the asymptotic analysis of the Green function starting with the Feynman path integral \cite{baranger1993,jalabert1990,gutzwiller2013, Fisher1981} gives an approximation of the S-matrix
\begin{align}
S_{b,a} (E) \approx \sum_{\beta: a\rightarrow b}^{\infty} A_{\beta} \exp \left(\frac{\mathrm{i}}{\hbar} \mathcal{S}_{\beta} \right)
 \label{eq:van Vleck formula}
\end{align}
in terms of a sum over all classical trajectories $\beta$ linking mode $a$ and $b$ at fixed energy $E$. The stability amplitude $A_\beta$ together with the phase factor depending on the classical action ${\mathcal{S}_\beta = \int_\beta \boldsymbol{p}\cdot \mathrm{d} \boldsymbol{q}}$ build up the contribution to $S_{b,a}$ of each classical trajectory $\beta$. 
For  time-reversal symmetric systems with  ${\mathcal{T}^2=1}$, such as the systems discussed here, the S-matrix is further symmetric: $S_{b,a}(E)=S_{a,b}(E)$ \cite{Beenakker1997} and therefore belongs to the Circular Orthogonal Ensemble (COE).

\subsection{Diagonal approximation}
\label{subsec:Diagonal approximation}
In a chaotic system, a quantity like
\begin{align}
    \langle S_{b,a} (E) S_{b,a}^* (E) \rangle =\Big \langle \sum_{\substack{\alpha:a\rightarrow b,\\ \beta: a \rightarrow b}} A_\alpha A^*_\beta e^{\frac{\mathrm{i}}{\hbar} (\mathcal{S}_\alpha-\mathcal{S}_\beta)} \Big \rangle 
\end{align}
with an energy average ${\langle \dots \rangle}$, will appear in a universal, system independent, form in accordance with the Bohigas-Gianonni-Schmit conjecture \cite{Bohigas1984}. For ${\mathcal{S}_\alpha - \mathcal{S}_\beta \gg \hbar}$, the highly oscillating  phase results in a vanishing contribution when averaging. In case of an action difference ${\mathcal{S}_\alpha - \mathcal{S}_\beta=0}$ and therefore $\alpha=\beta$ (excluding systems with exact discrete symmetries \cite{Baranger1996}), the so-called diagonal approximation gives the leading contribution in orders of $1/N$ \cite{Richter2002, Kuipers_2009, Novaes_2corr, Ericson2016}:
\begin{align}
    \langle S_{b,a} (E) S_{b,a}^* (E) \rangle = \frac{1}{N} + \mathcal{O} \left( \frac{1}{N^2}\right).
\end{align}
Higher orders of this correlator, like the contribution by the Richter/Sieber pairs depicted in Fig.~\ref{fig:quadruplets} (a), will be discussed later.
\subsection{Semiclassical calculation of energy dependent correlators}
\label{subsec:Semiclassical calculation of energy dependent correlators}
\begin{figure}[ht]
\centering
 \begin{overpic}[width=0.99 \linewidth]{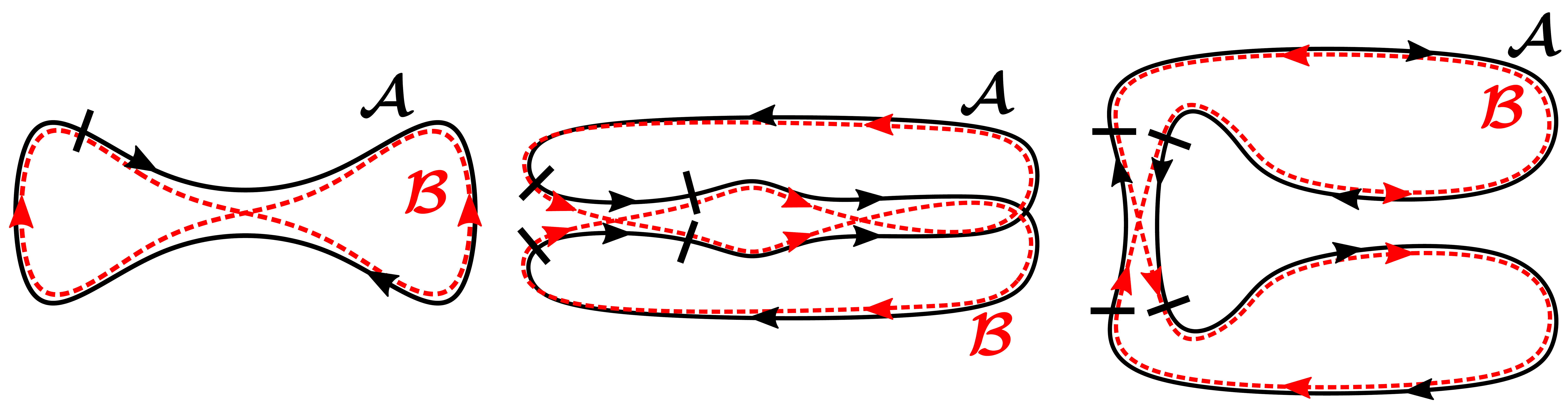}
   \put (0,25) {\large \textbf{(a)}}
   \put (34,25) {\large \textbf{(b)}}
   \put (71,25) {\large \textbf{(c)}}
 \end{overpic}

\caption{Constructing semiclassical contributions to S-matrix correlators. Each contribution to $C_2$ is constructed by cutting once through a periodic orbit pair, \textit{e.g.} the pair sketched in (a). The quadruplets for the four-point correlators are generated by cutting out one \mbox{2-encounter}. For the $D^\times$ correlator this 2-encounter has to be parallel in $\mathcal{A}$ and $\mathcal{B}$ as depicted in (b). (c): Instead using an antiparallel 2-encounter yields the trajectory pairs for $C^=$ and $B^=$  \cite{M_ller_2007}.}
\label{fig:quadruplets}
\end{figure}

The correlators of interest,
\begin{align}
    C^= (\Delta) &:= \Big\langle \vert S_{b,a} (E_+) \vert^2 \; \vert S_{d,c} (E_-) \vert^2  \Big\rangle,  \label{eq: definition C parallel}\\
    D^\times (\Delta) &:= \langle S_{b,a} (E_+) S_{d,c} (E_-) S^*_{d,a}   (E_-) S^*_{b,c} (E_+) \rangle, \label{eq: definition D antiparallel}\\
    B^=(\Delta) &:= \langle S_{b,a} (E_+) S_{d,c} (E_+) S^*_{b,a}  (E_-) S^*_{d,c} (E_-) \rangle, \label{eq: definition B parallel}
\end{align} 
with distinct modes $a,b,c$ and $d$ are depending, due to the average, only on the energy difference $\Delta$ defined by ${E_{\pm}= \bar{E} \pm \Delta=\bar{E} \pm \frac{\eta \hbar}{2 \tau_D}}$, where we introduced the dwell time $\tau_D$. For example, coherent contributions to $D^\times$ can be made explicit when inserting Eq.~(\ref{eq:van Vleck formula}) in Eq.~(\ref{eq: definition D antiparallel}) 
\begin{align}
    D^\times (\Delta)=  
    \Big\langle \sum_{\substack{\alpha:a\rightarrow b,\\ \beta: a \rightarrow d}} \sum_{\substack{\gamma:c\rightarrow d,\\ \delta:c \rightarrow b}} A_\alpha A^*_\beta A_\gamma A^*_\delta  e^{\frac{\mathrm{i}}{\hbar} \Delta \mathcal{S} } \Big\rangle.
\end{align} 
with ${\Delta \mathcal{S}:=\mathcal{S}_\alpha \left( E_+\right)-\mathcal{S}_\beta \left( E_-\right)+\mathcal{S}_\gamma \left( E_-\right)-\mathcal{S}_\delta \left( E_+\right)}$. For $\Delta \mathcal{S}$ of order $\hbar$, the contribution to $D^\times$ for this quadruplet ${(\alpha, \beta, \gamma, \delta)}$ can be identified by a diagrammatic rule \cite{M_ller_2007}. Each quadruplet can be generated by following the recipe: The periodic orbit pairs $\mathcal{A}$ and $\mathcal{B}$ as depicted in Fig.~\ref{fig:quadruplets} (b) and (c) are building the basis. Removing one 2-encounter (thick lines in Fig.~\ref{fig:quadruplets}) of all periodic orbit pairs, all relevant quadruplets contributing to the correlators are generated. Depending on the correlator, the removed 2-encounter has to be either parallel in $\mathcal{A}$ and  $\mathcal{B}$ (in case of $D^\times$) or antiparallel in $\mathcal{A}$ and $\mathcal{B}$ (in case of $C^=$ and $B^=$) as shown in Fig.~\ref{fig:quadruplets}(b) and (c). 
For $\Delta=0$, the correlators are known \cite{Novaes_2corr}:
\begin{align}
\begin{split}
D^\times (\Delta=0)&= -\frac{1}{N(N+1)(N+3)} ,\\
B^=(\Delta=0) &= C^= (\Delta=0)= \frac{N+2}{N(N+1)(N+3)}.
\end{split}\label{eq: D cross, C and B parallel correlator Delta zero}    
\end{align}

Besides cutting the periodic orbit pairs just once as illustrated in Fig.~\ref{fig:quadruplets} (a), the same steps are needed to evaluate the two-point correlator semiclassically~\cite{Kuipers_2009,Ericson2016} 
\begin{align} \label{eq: 2.correlator semiclassic}
\begin{split}
C_2 &(\Delta):= \langle S_{b,a} (E_+) S_{b,a}^* (E_-) \rangle \\
&=\frac{1}{N (1-\mathrm{i} \eta)} - \frac{1-2 \mathrm{i} \eta}{N^2 (1- \mathrm{i} \eta)^3} \\
&+ \frac{\mathrm{i}+4\eta -8 \mathrm{i} \eta^2 }{N^3(\mathrm{i}+ \eta)^5} +\frac{\mathrm{i}+6 \eta-12\mathrm{i}\eta^2-48\eta^3}{N^4 (\mathrm{i}+ \eta)^7} + \mathcal{O} \left(\frac{1}{N^5} \right). 
\end{split}
\end{align}

Using this powerful method for energy-independent transport moments the Bohigas-Gianonni-Schmit conjecture was proven in Refs.~\cite{Berkolaiko2012, Berkolaiko2013}.

\section{Experiment}
\label{sec:Experiment}
\subsection{Quantum graphs as model systems for quantum chaos}
\label{subsec:Quantum graphs as model systems for quantum chaos}
Quantum graphs~\cite{Kottos1997,Kottos1999,Pakonski2001,Texier2001} have been used for two decades to study the features of quantum chaos~\cite{Stoeckmann1999,Haake2001} in closed and open quantum systems. They exhibit several properties which make them most suitable for studies within the field of quantum chaos. (i) The spectral properties of closed quantum graphs with incommensurable bond lengths were proven in Ref.~\cite{Gnutzmann2004} to coincide with those of random matrices from the Gaussian ensemble~\cite{Mehta1990} of the same universality class~\cite{Gnutzmann2004}, in accordance with the Bohigas-Gianonni-Schmit conjecture for chaotic systems~\cite{Berry1979,Casati1980,Bohigas1984}. (ii) The semiclassical trace formula, which expresses the fluctuating part of the spectral density in terms of a sum over the classical periodic orbits, is exact~\cite{Kottos1999}. (iii) It was demonstrated in Refs.~\cite{Pluhar2013,Pluhar2013a,Pluhar2014} that the correlation functions of the $S$-matrix elements of open quantum graphs with a classically chaotic scattering dynamics coincide with the corresponding RMT results. It is also worth to mention that quantum graphs belonging to the orthogonal, the unitary and the symplectic universality class have been realized experimentally with microwave networks consisting of coaxial cables connected by joints~\cite{Hul2004,Lawniczak2010,Allgaier2014,Bialous2016,Rehemanjiang2016,Lu2020}. 

\subsection{Setup for energy dependent correlations}\label{subsec:Setup for energy dependent correlations}
\begin{figure}[ht]
        \centering
        \includegraphics[width=0.6\linewidth]{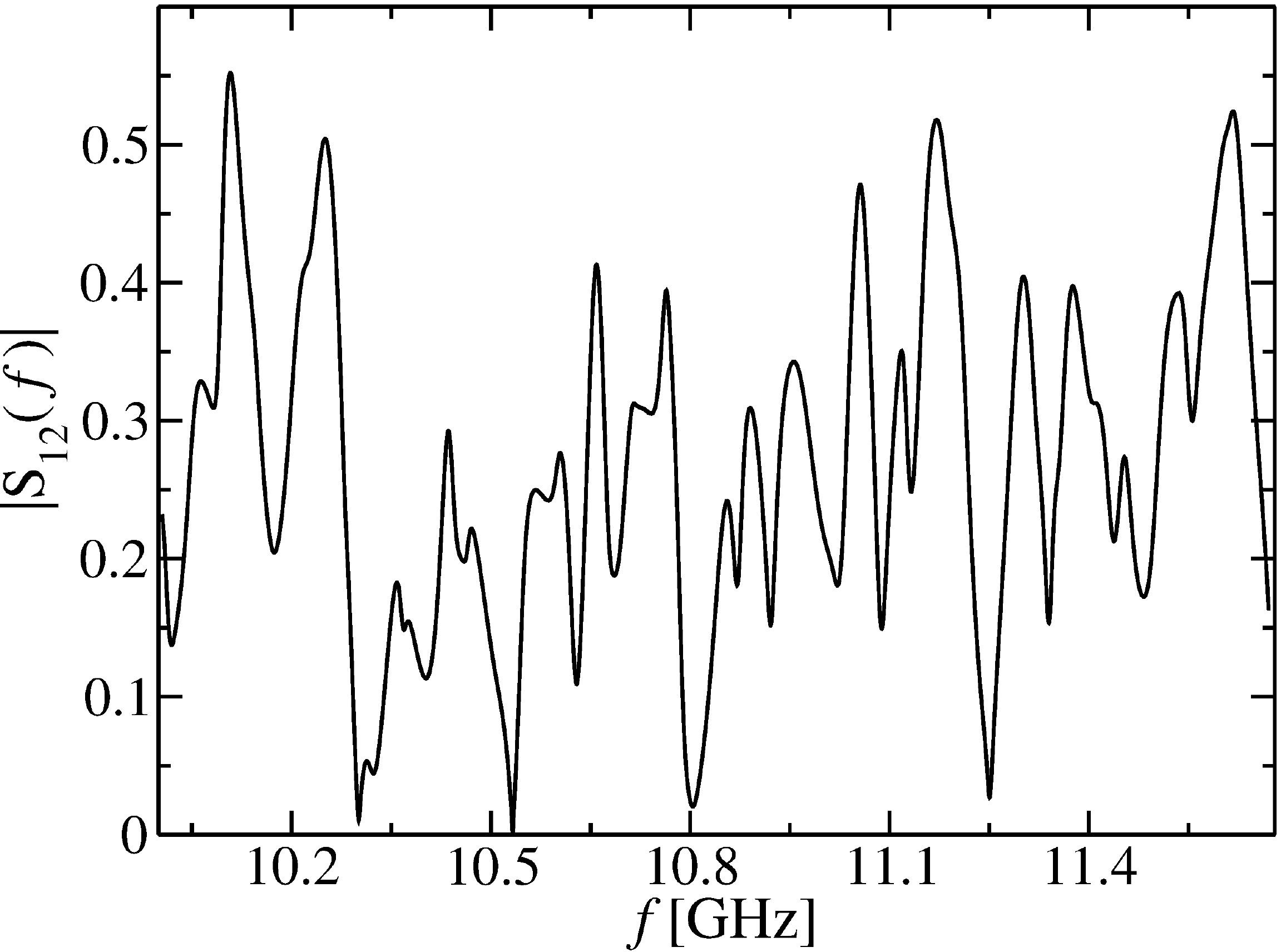}
        \caption{Modulus of the S-matrix element $S_{12}$ in the frequency interval $f\in[10,11.6]$~GHz where the correlation functions were analyzed.}
        \label{fig: abs(S) experiment new}
        \end{figure}

\begin{figure}[ht]
        \centering
        \includegraphics[width=0.48\linewidth]{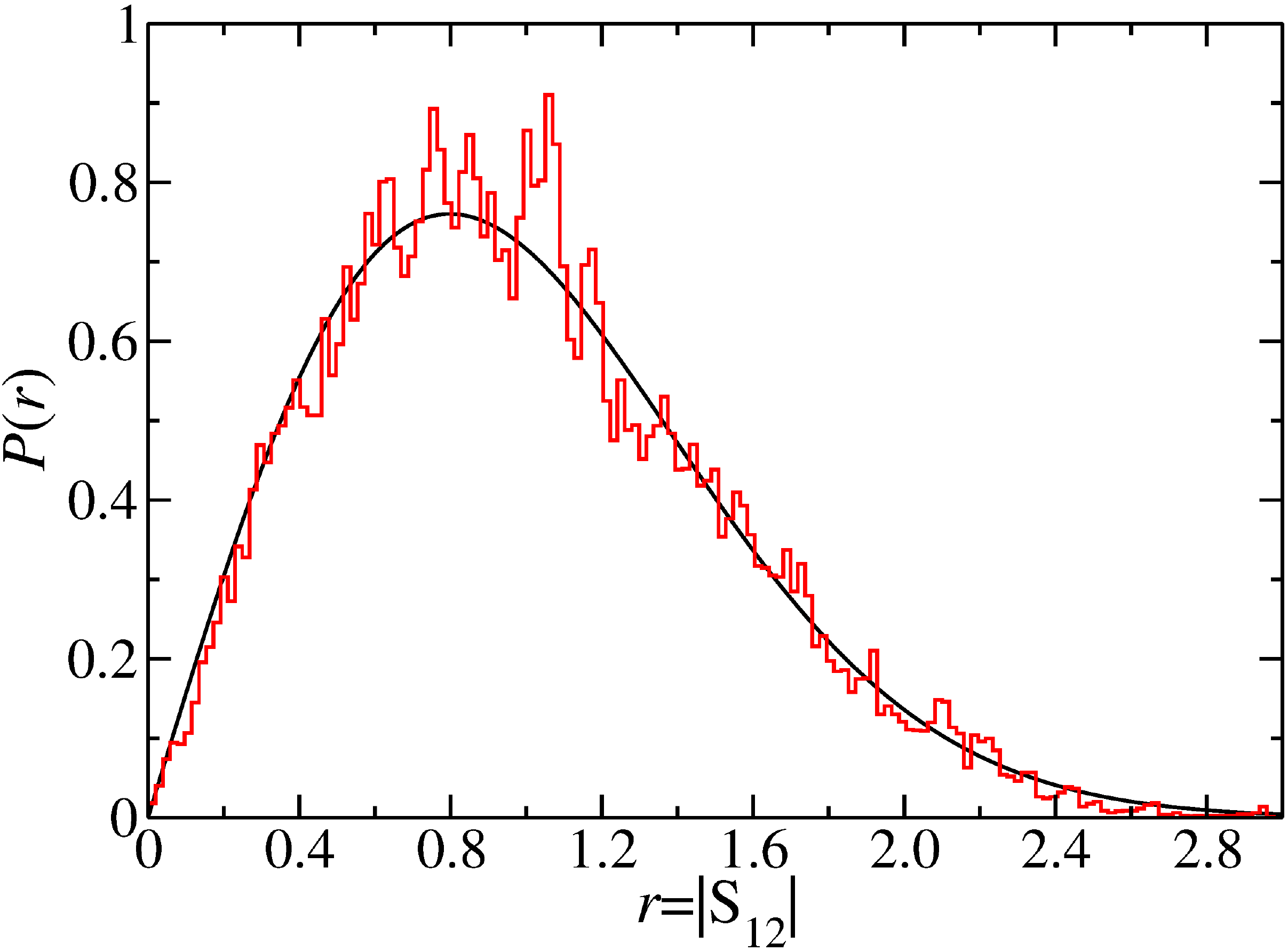}
        \includegraphics[width=0.5\linewidth]{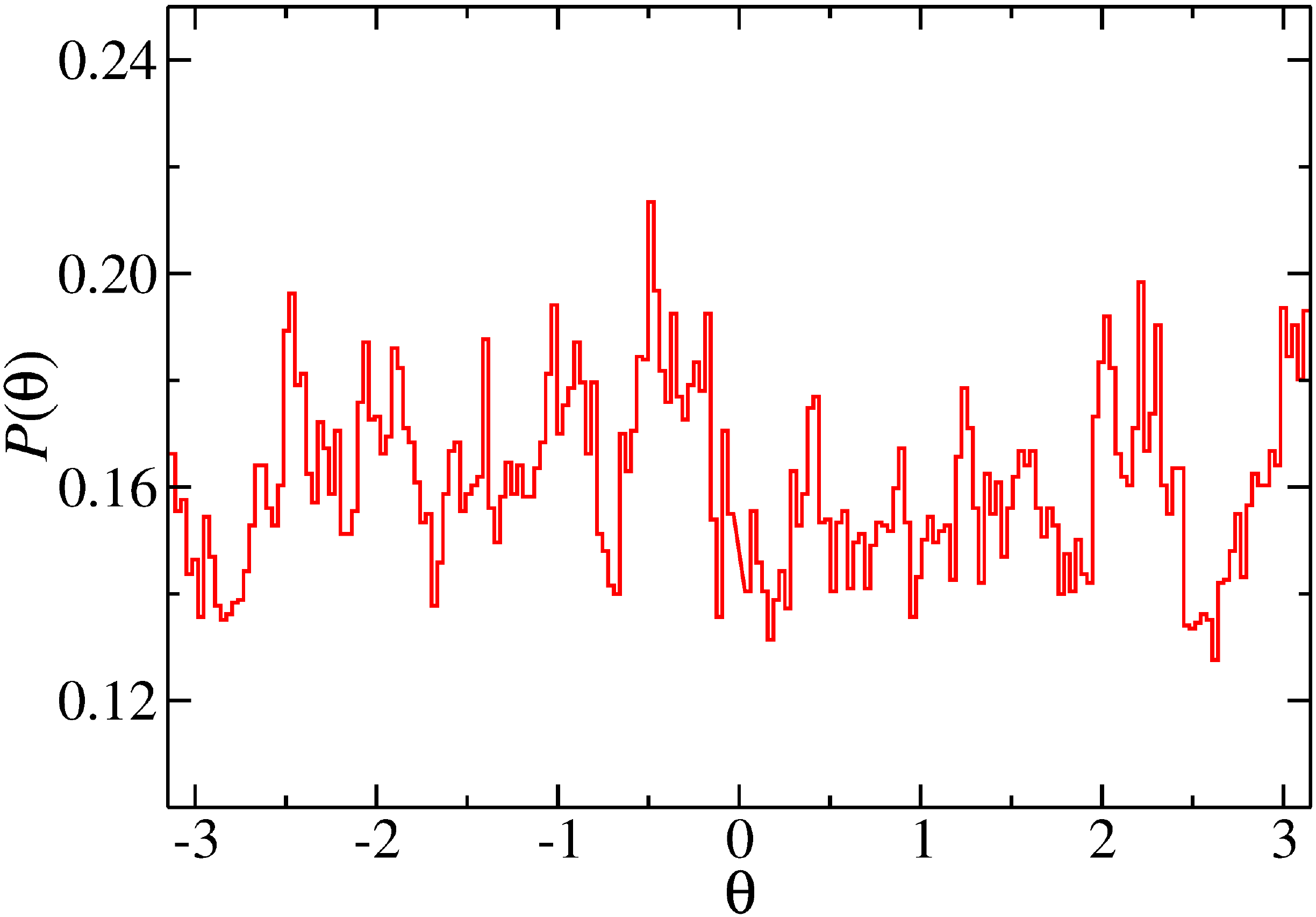}
        \caption{Distribution of the experimental S-matrix element $S_{12}$ (red curves). As expected in the Ericson region of strongly overlapping resonances, the distributions of the modulus (left) and phase (right) are bivariate Gaussian (black curve) and uniform, as predicted by RMT (black line).}
        \label{Distr_S12}
        \end{figure}

We constructed a microwave network preserving time-reversal invariance and, thus, belonging to the orthogonal universality class, which consists of 19 coaxial cables and 14 T joints corresponding to the bonds and vertices of valency 3 in the associated quantum graph, respectively. Five additional networks were generated by interchanging two bonds which are connected to distinct vertices. Here, the lengths of the bonds were chosen incommensurable to attain a 'chaotic' quantum graph. Antennas, i.e., leads, were attached to four of them and the $4\times 4$ S-matrix was measured~\cite{Lu2020}. Thus, the number of open channels equals $N=4$ in an ideal microwave network. The experimental correlation functions were obtained by averaging over 30 correlation functions obtained for the 6 realizations in five frequency ranges of about 0.5~GHz in the interval $f\in[10,11.6]$~GHz. A part of a measured transmission spectrum is shown in Fig.~\ref{fig: abs(S) experiment new}.
To check, whether the experimental data comply with chaotic scattering with perfect coupling to the continuum we first analyzed the distribution of the modulus and phase of the measured $S_{ab}$. The modulus is bivariate Gaussian and the phases are uniformly distributed, as is expected in the Ericson region~\cite{Agassi1975, Kumar2013}. This is illustrated in Fig.~\ref{Distr_S12}. 

\begin{figure}[ht]
        \centering
        \includegraphics[width=0.49 \linewidth]{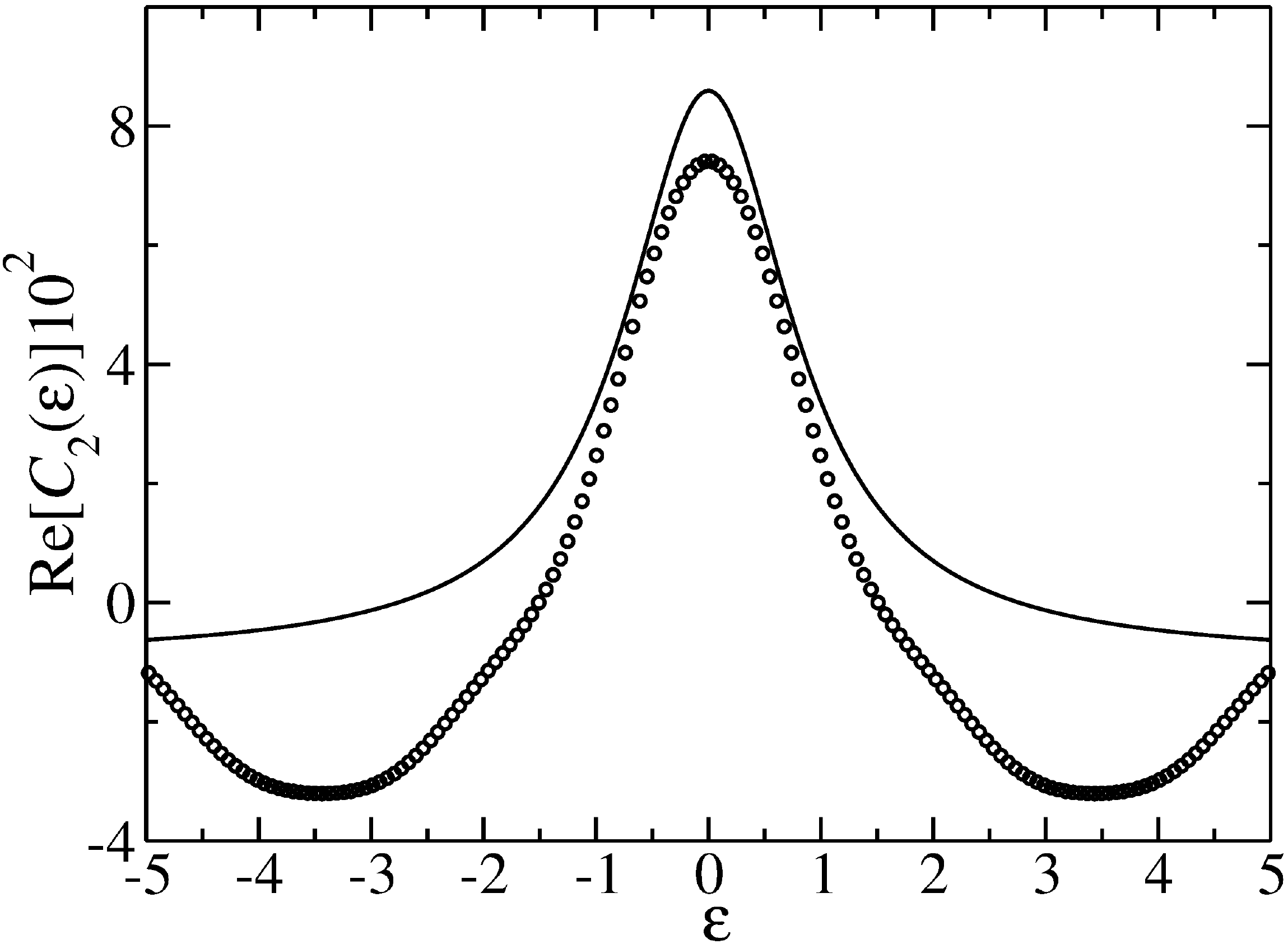}
        \includegraphics[width=0.49 
        \linewidth]{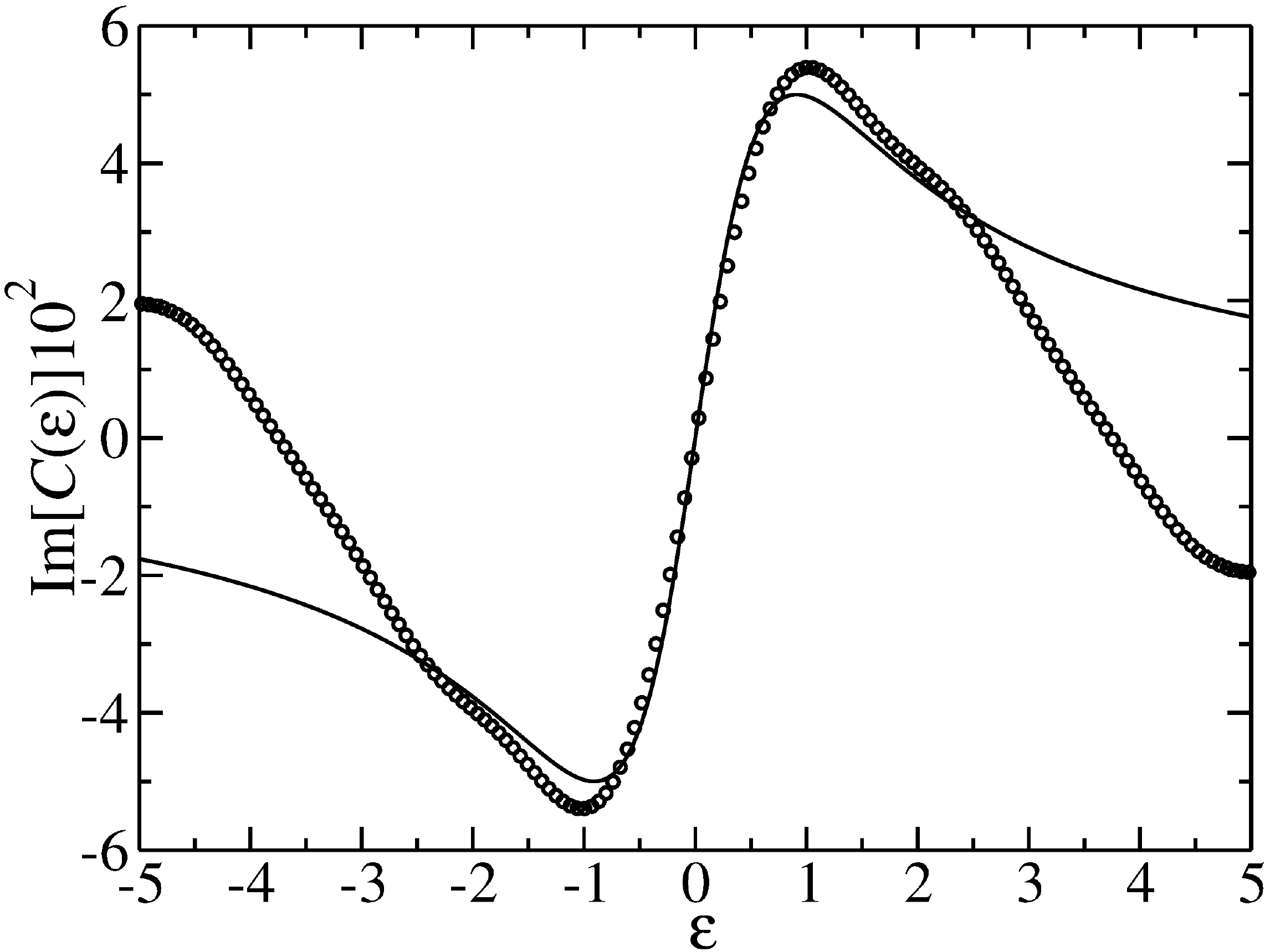}
	\caption{Experimental results (black dots) for the real (left) and imaginary (right) parts of the two-point correlation function. Adjusting Eq. (\ref{eq: 2.correlator semiclassic}) to the curve (black lines) yields $N\simeq 9-10$ open channels and $\tau_{QG}\approx 0.91\tau_D$. The dimensionless quantity $\epsilon=\Delta/D$ represents energy rescaled by the resonance spacing $D$ of the microwave graph.}
        \label{fig:2.correlator Experiment}
\end{figure}

Note, that absorption is unavoidable in the experiments, that is, realization of an ideal microwave network is unfeasible. This leads to a further opening of the system which may be accounted for, e.g., by introducing additional open channels. In order to obtain an estimate for the effective number of open channels we adjusted Eq.~(\ref{eq: 2.correlator semiclassic}) to the experimental two-point correlator, where we varied $N$ and also the scale of $\epsilon$, i.e., of $\tau_D$, by introducing a fit parameter $\tau_{QG}$, $\varepsilon\to\varepsilon\frac{\tau_D}{\tau_{QG}}$, of which the value is restricted to $0.9\leq\tau_{QG}\leq 1.1$, to account for a possible error in the determination of the total optical length $\mathcal{L}$ of the network and thus of the average resonance spacing $D=c/(2\mathcal{L}$), where $c$ denotes the velocity of light. Best agreement between the experimental results and Eq.~(\ref{eq: 2.correlator semiclassic}) was obtained for $N\simeq 9-10$ open channels and $\tau_{QG}\approx 0.91\tau_D$, shown in Fig.~\ref{fig:2.correlator Experiment}.

\section{Numerical approaches}
\label{sec:Numerical approaches}
\subsection{Tight-binding model}
\label{subsec:Tight-binding model}

\begin{figure}
\centering
\includegraphics[width=0.2 \textwidth]{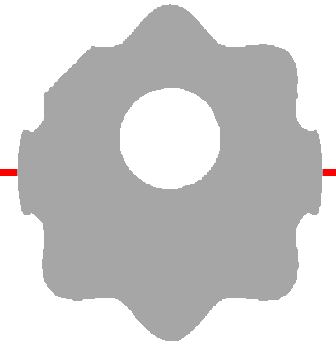}
\caption{Desymmetrized cavity for chaotic scattering. The boundary of the tight-binding setup is undulatory. Inside, the circular obstacle improves ergodic behaviour. The two metallic contacts, with a total number of open modes $N=10$ or $N=4$, are shown in red.}
\label{fig: system}
\end{figure}
\begin{figure}[ht]
\centering
\includegraphics[width=0.49 \linewidth]{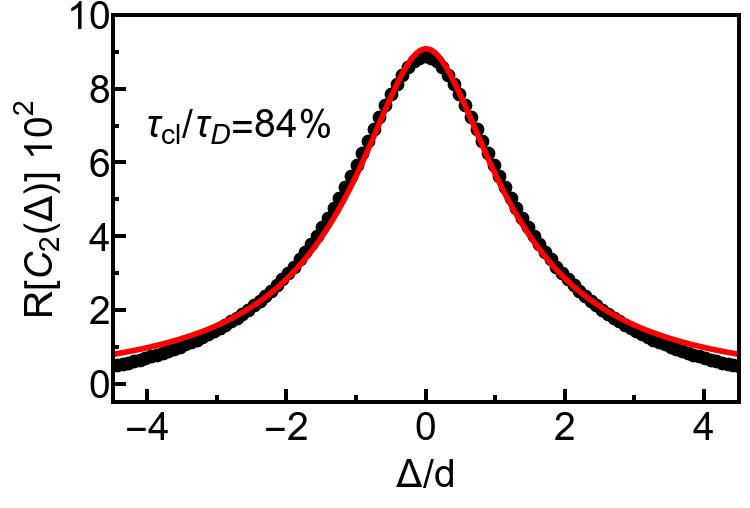}
\centering
\includegraphics[width=0.49 \linewidth]{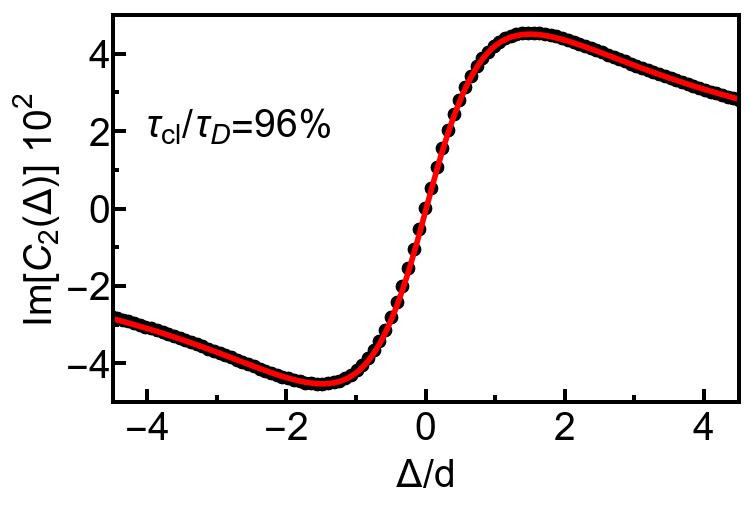}
\caption{The system with $N=10$ (black dots) exhibits the form of the real (left panel) and imaginary (right) part of the two-point correlator $C_2 \left( \Delta \right)$ given in Eq.~(\ref{eq: 2.correlator semiclassic}) (black lines). The agreement of the fitted dwell time $\tau_D$ with the classical one, ${\tau_{cl}/v_g= 52}$, is clearly observable.}
\label{fig:2.correlator KWANT}
\end{figure}

The tight-binding model, implemented with the Kwant code~\cite{Groth_2014} and the Hamiltonian ${\hat{H}=\frac{\hbar^2}{2 m_e} k^2}$ with the free electron mass $m_{e}$. 
The two-dimensional system is connected to two metallic contacts. The scattering problem between these two channels is solved based on the wave function approach \cite{Groth_2014}. To introduce chaotic scattering into the system, a billiard-shaped setup  seems to be reasonable. Reaching true ergodic scattering in this setup is strikingly difficult. The system's shape, which nearly exhibits ergodicity, is illustrated in Fig.~\ref{fig: system}, with the two leads marked red. The elliptical system requires an additional disk-like obstacle and uncorrelated weak disorder. The  mean free path $l_{el}$ 
resulting from scattering at the disorder is fulfilling the conditions of weak disorder (${k_F \cdot l_{el} \gg 1}$), ballistic scattering (${\lambda_F <L<l_{el}}$) and of  the non-localized regime (${L<N\cdot l_{el}\approx \xi_{loc}}$) with $L$ the typical system length. In the following this setup is investigated for total number of open modes ${N=10}$ and ${N=4}$. With the area $A$, total width of all leads $\mathcal{C}$ and group velocity $v_g$, the classical dwell time is given by ${\tau_{cl}=\frac{\pi A}{\mathcal{C} v_g}}$ and the mean level distance by $d=\frac{\hbar^2}{m_{e} A}$. For the two adjustments we expect a dwell time of  around  ${\tau_{cl}/v_g= 52}$ (for ${N=10}$) and  ${\tau_{cl}/v_g= 156}$ (for ${N=4}$). The statistics of ${\vert S_{b,a} (E) \vert}$, ${\arg (S_{b,a} (E) )}$ are bivariate Gaussian, respectively, uniformly distributed. Adapting the two-point correlator ${C_2 (\Delta)}$ of Eq.~(\ref{eq: 2.correlator semiclassic}) to the simulation with $N=10$ by fitting the dwell time $\tau_D$ and a prefactor, indicates the desired chaotic scattering as shown in Fig.~\ref{fig:2.correlator KWANT}. To this end, averages over disorder configurations, sets of channel combinations ${(a,b,c,d)}$ and mean energies $\bar{E}$ are necessary. 

\subsection{Quantum graph with absorption}
\label{subsec:Quantum graph with absorption}
In order to quantify the effect of absorption on the $S$-matrix correlations, we analyzed the properties of the corresponding quantum graph with and without absorption, where we chose Neumann boundary conditions at the vertices~\cite{Kottos1999}. The $N\times N$ dimensional S-matrix of a quantum graph with $N$ attached leads can be written in the form~\cite{Kottos1999} 
\begin{align}
\begin{split}
S_{b,a}(f) &= \delta_{b,a} - 2\pi \mathrm{i}\left[\hat W^\dagger\left(\hat H(k)+\mathrm{i}\pi\hat W\hat W^\dagger\right)^{-1}\hat W\right]_{b,a},\, \\
a,b&=1,\dots,N
\end{split}
\label{eqn:Sab}
\end{align}
with the wave number $k$ being related to frequency via $k=2\pi f/c$. Furthermore, $\hat H(k)$ is the Hamiltonian of the closed quantum graph and $\hat W$ is a rectangular matrix of dimension $\mathcal{V}\times N$ accounting for the coupling of the leads to the $\mathcal{V}$ vertices, i.e., $W_{ij},\, i=1,\dots,\mathcal{V},\, j=1,\dots,N$ equals unity if lead $j$ is coupled to vertex $i$ and zero otherwise. We accounted for absorption by adding a small imaginary part to the wavenumber $k$. These numerical computations essentially confirmed our experimental results. We furthermore performed RMT simulations for quantum chaotic scattering systems using the Heidelberg approach~\cite{Mahaux1966,Dietz2010}.

\subsection{The Heidelberg approach within RMT}
\label{subsec:The Heidelberg approach within RMT}
In the Heidelberg approach the graph Hamiltonian $\hat{H} (k)$ is replaced by $\hat{H} (k)=k \II -\mathcal{H}$, where $\mathcal{H}$ is a $M\times M$-dimensional random matrix from the Gaussian orthogonal ensemble in our case, and $\hat W$ is replaced by a matrix with real, Gaussian distributed entries with zero mean. This is only possible when restricting the length of the frequency intervals such that the average resonance parameters are approximately constant~\cite{Dietz2010,Lu2020}. The average of the $S$ matrix over frequency is diagonal, ${\langle S_{b,a}\rangle=\langle S_{a,a}\rangle \delta_{a,b}}$, implying~\cite{Mahaux1966} that $\sum_{\mu = 1}^M W_{e \mu} W_{e^\prime \mu}=M v_{e}^2 \delta_{e,e^\prime}$. The parameter $v^2_{e}$ measures the average strength of the coupling of the modes excited in the coaxial cables to the lead $e$. Generally, it is related to the transmission coefficients $ T_{e} = 1 - |\left\langle{S_{e,e}}\right\rangle |^2 $ via $T_{e} = \frac{4 \pi^2 v^2_{e} / \tilde D}{(1 + \pi^2 v^2_{e} / \Tilde{D})^2}$ with $\tilde D=\sqrt{\frac{2}{M}\langle H_{\rm\mu\mu}^2\rangle}\frac{\pi}{M}$ denoting the mean resonance spacing. We verified that coupling to the antennas is perfect in the experiments, that is $T_{e}\simeq 1$ for $e=1,\dots ,N$ by comparing the distributions of the S-matrix elements to the RMT predictions [Fig.~\ref{Distr_S12}]. The RMT simulations were performed for ($200\times 200$)-dimensional random matrices and an ensemble of 300 S-matrices was generated. In order to model absorption~\cite{Dietz2010} we added scattering channels to the four open channels, where the transmission coefficients were set equal to unity. We also chose $N=4$ and simulated absorption by $\Lambda =50$ fictitious channels~\cite{Dietz2010} with equal transmission coefficients $T_f\ll 1$.

\section{Results and discussion}
\label{sec:Results and discussion}

\begin{figure}[ht]
\centering
 \begin{overpic}[width=0.49    \linewidth]{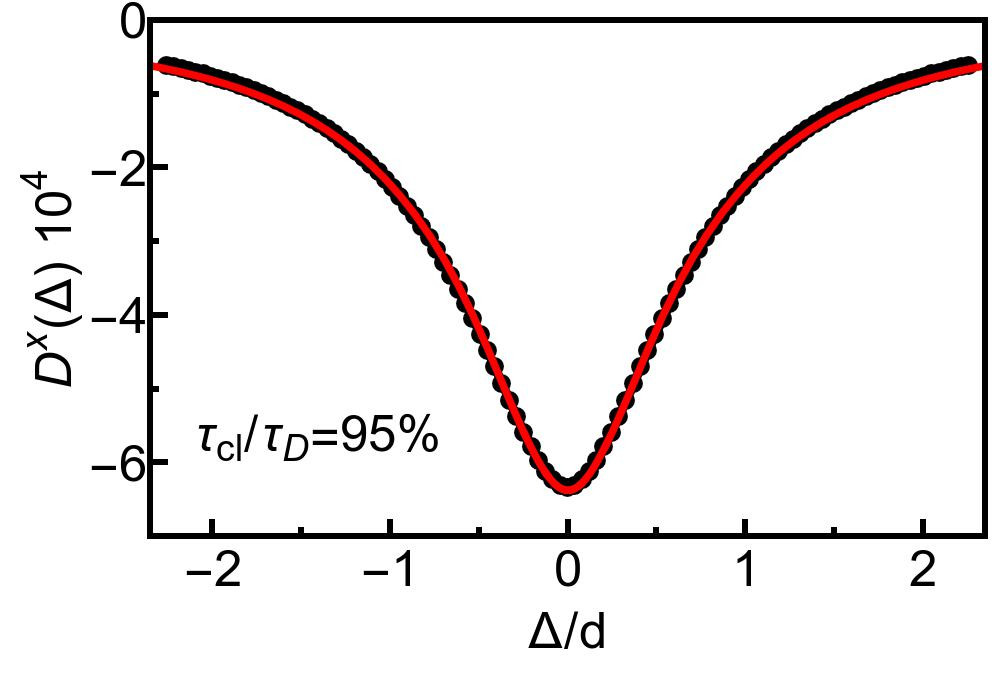}
   \put (-2,65) {\large \textbf{(a)}}
 \end{overpic}
 \centering
 \begin{overpic}[width=0.49   \linewidth]{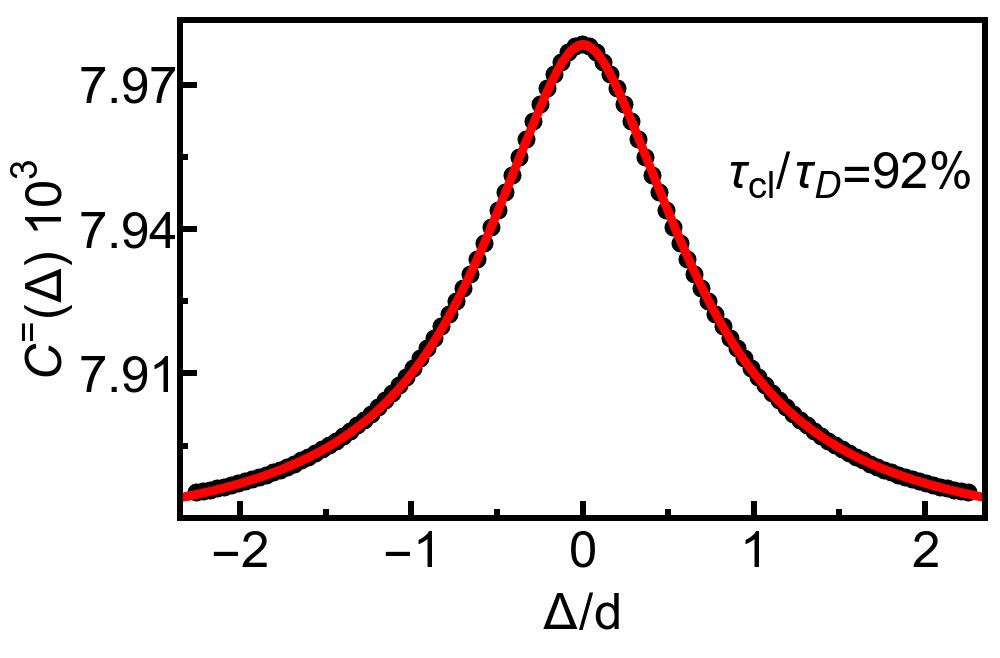}
   \put (-2,65) {\large \textbf{(b)}}
 \end{overpic}\\
 \centering
 \begin{overpic}[width=0.49    \linewidth]{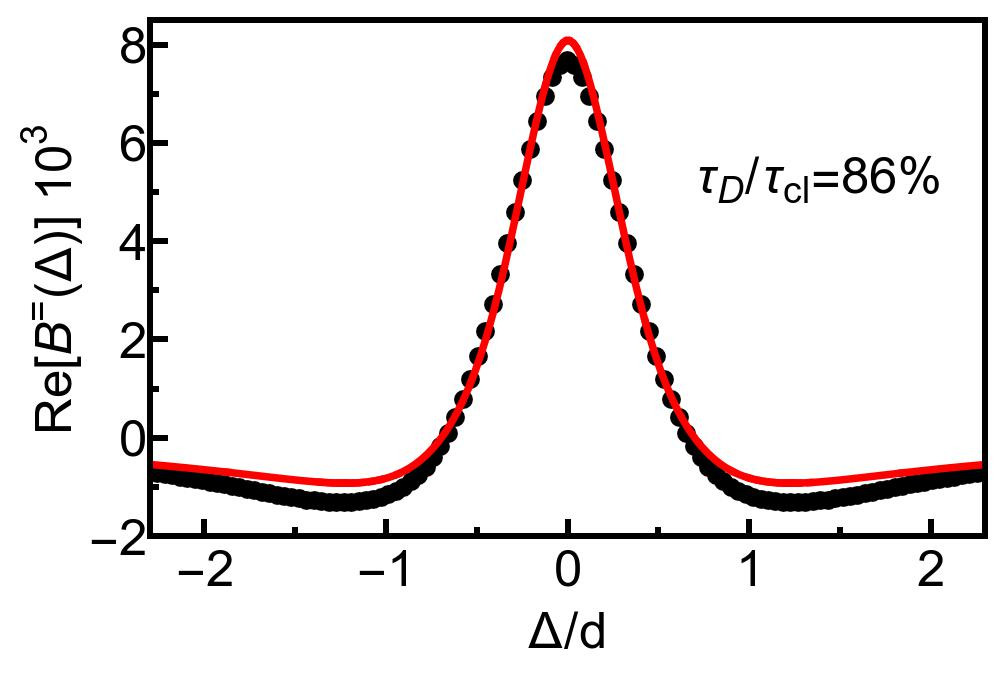}
   \put (-2,65) {\large \textbf{(c)}}
 \end{overpic}
 \centering
 \begin{overpic}[width=0.49   \linewidth]{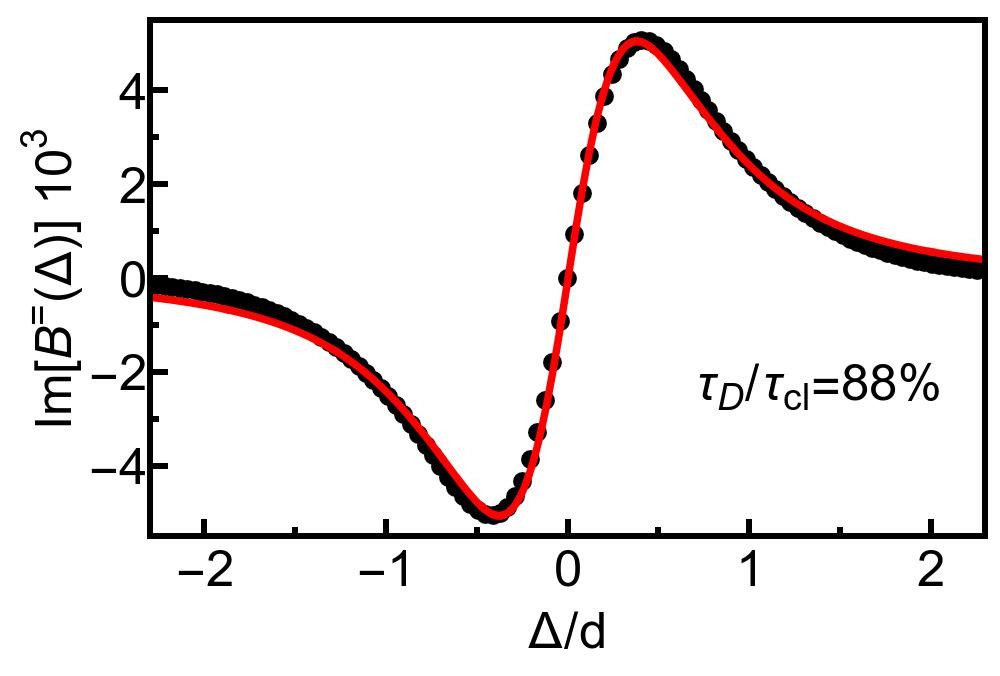}
   \put (-2,65) {\large \textbf{(d)}}
 \end{overpic}

\caption{Four-point correlators: Semiclassics versus numerics. For ${N=10}$ the numerical simulation  of the four-point correlators $D^\times$, $C^=$ and $B^=$ (black dots) coincide with fits of Eqs.~(\ref{eq: D cross correlator semiclassic})-(\ref{eq: B parallel correlator semiclassic}) (red lines). Almost no changes appear when taking higher orders in $1/N$ into account.}
\label{fig:four-point corr for N10 KWANT}
\end{figure}

\begin{figure}[ht]
\centering
 \begin{overpic}[width=0.49 \linewidth]{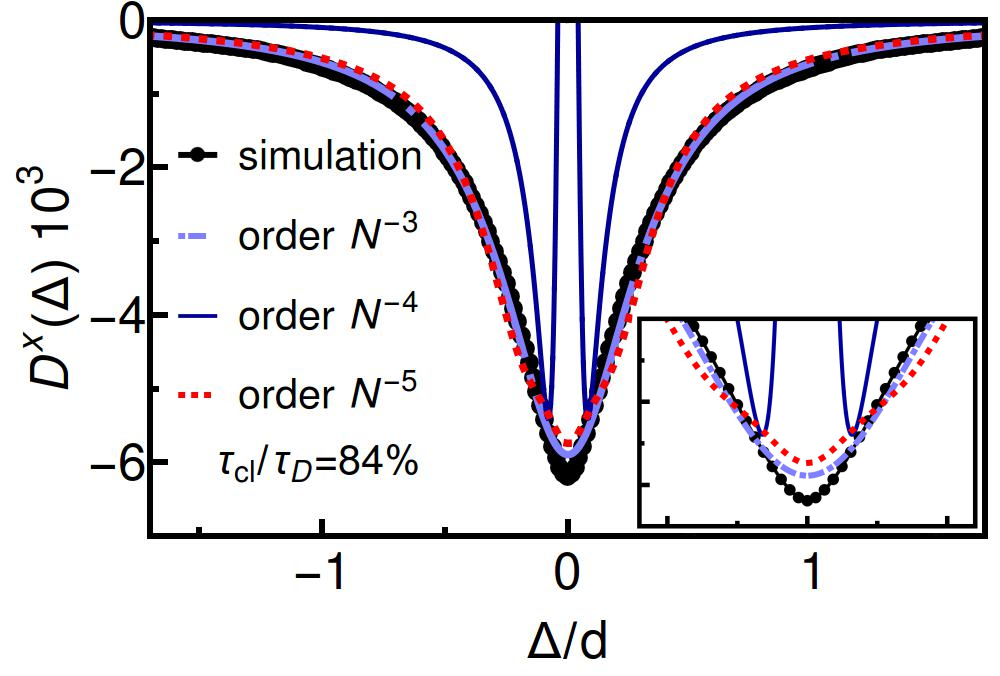}
   \put (-2,65) {\large \textbf{(a)}}
 \end{overpic}
 \centering
 \begin{overpic}[width=0.49   \linewidth]{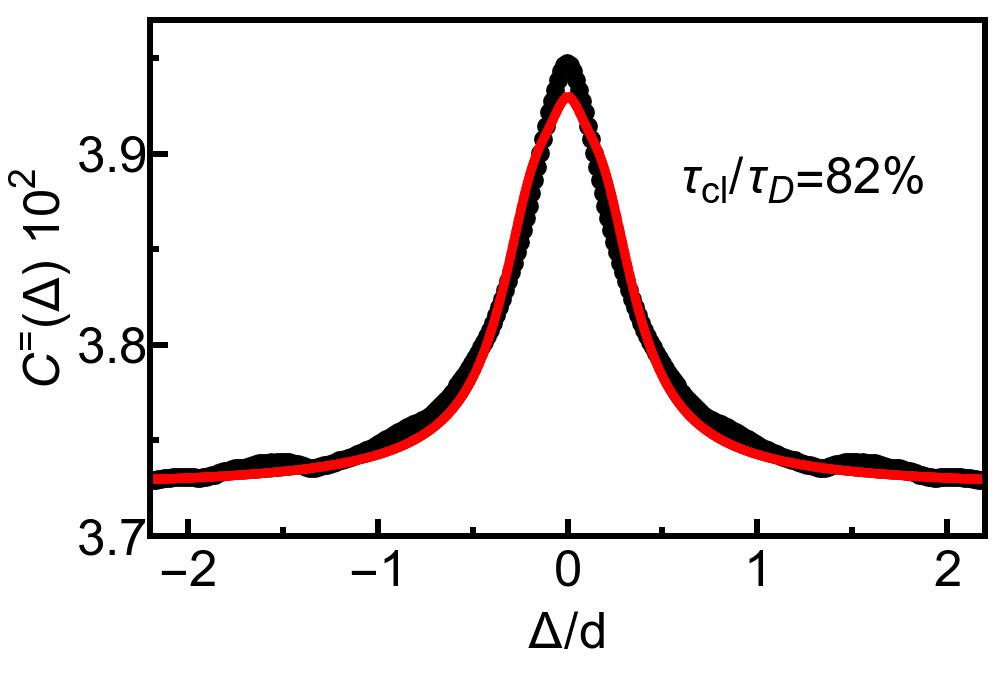}
   \put (-2,65) {\large \textbf{(b)}}
 \end{overpic}\\
 \centering
 \begin{overpic}[width=0.49    \linewidth]{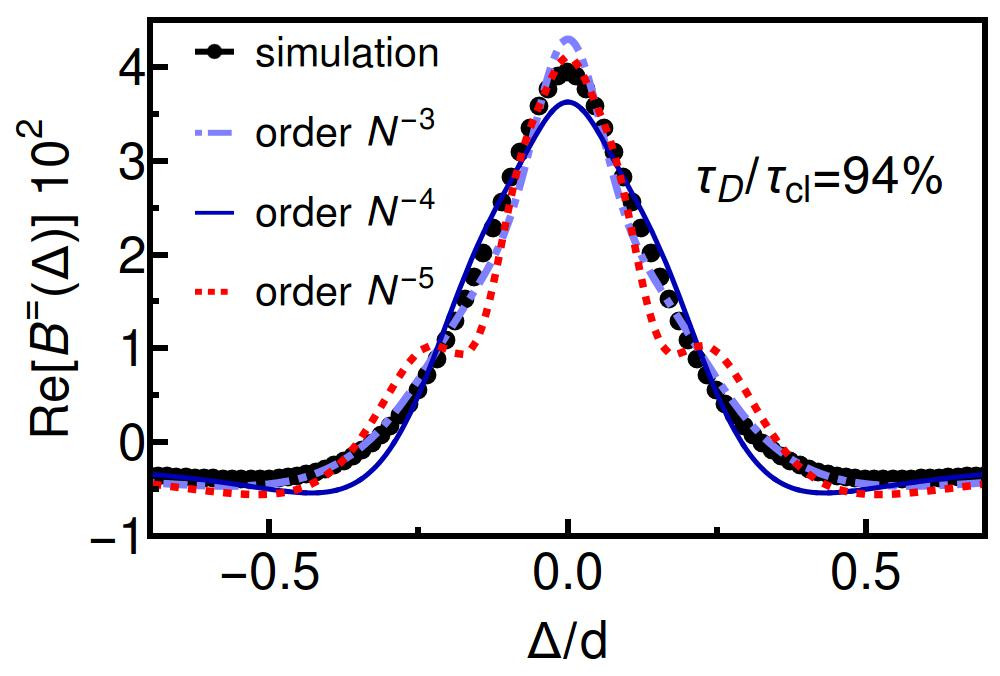}
   \put (-2,65) {\large \textbf{(c)}}
 \end{overpic}
 \centering
 \begin{overpic}[width=0.49   \linewidth]{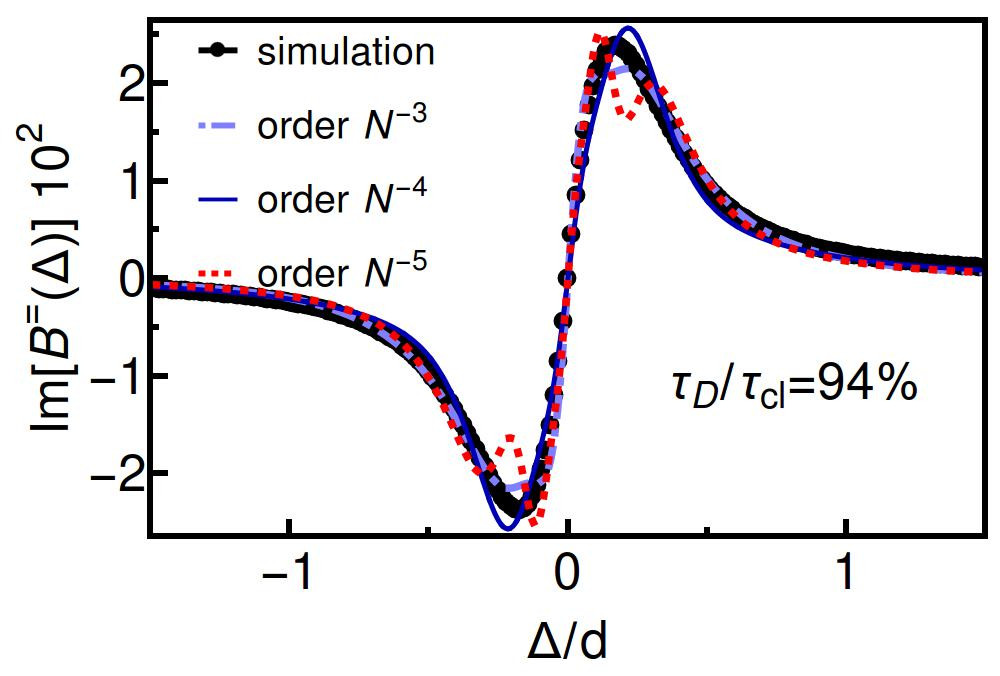}
   \put (-2,65) {\large \textbf{(d)}}
 \end{overpic}

\caption{Slow convergence of semiclassics for $D^\times$ for ${N=4}$, (a) (lines), due to large cancellations in Eq.~(\ref{eq: D cross correlator semiclassic}). Accordingly, $C^=$ of Eq.~(\ref{eq: C parallel correlator semiclassic}) (red line) oscillates, however in  (b) just order $N^{-6}$ is shown. The agreement to numerics (black dots) is partly still present. In comparison, Eq.~(\ref{eq: B parallel correlator semiclassic}) of the $B^=$ correlator (lines) converges faster and coincides with simulations (black dots) for all orders (c), (d).}
\label{fig:four-point corr for N4 KWANT}
\end{figure}

The semiclassical analysis described in Section \ref{subsec:Semiclassical calculation of energy dependent correlators} for the energy dependent four-point correlators can be written in orders of $1/N$. The simplest quadruplet accounting for the real correlator $D^\times$ is illustrated in Fig.~\ref{fig:quadruplets} (b). Taking more complex quadruplets into account results in
\begin{align}
D^\times &(\Delta)=  - \frac{1}{N^3 (1+\eta^2)} + \frac{2 \eta^4 + 10 \eta^2 +4}{N^4 (1+\eta^2)^3} \nonumber \\ 
\begin{split} 
&- \frac{8\eta^8 +30 \eta^6 +145 \eta^4 +56 \eta^2 +13}{N^5 (1+\eta^2)^5} +  \frac{28 \eta^{12} +190 \eta^{10}}{N^6 (1+\eta^2)^7}\\ 
&+ \frac{196 \eta^8 +2832 \eta^6 +392 \eta^4 +258 \eta^2+40}{N^6 (1+\eta^2)^7} - \mathcal{O} \left(\frac{1}{N^7} \right).
\end{split}
\label{eq: D cross correlator semiclassic}
\end{align} 
Equivalent thereto, cutting the trajectory pairs of Fig.~\ref{fig:quadruplets} (c) at the thick markers, the trajectory pairs yield the first order of the real correlator $C^=$ and the complex correlator $B^=$ correlator. Following this procedure we obtain the contributions 
\begin{align}
C^=& (\Delta) = \frac{1}{(N+1)^2} + \frac{2}{N^4 (1+\eta^2)}-\frac{6\eta^4 +24\eta^2 +10}{N^5 (1+\eta^2)^3} \nonumber \\
\begin{split}
& + \frac{22 \eta^8 +96 \eta^6 +354 \eta^4 +156 \eta^2+36}{N^6 (1+\eta^2)^5} 
 +\mathcal{O} \left(\frac{1}{N^7} \right)
\end{split}
\label{eq: C parallel correlator semiclassic}
\end{align}
and
\begin{align}
B^=(\Delta) &=-\frac{1}{N^2 (\mathrm{i}+\eta)^2}- \frac{4\mathrm{i} \eta-2}{N^3 (\mathrm{i}+\eta)^4} + \frac{32 \eta^2+20\mathrm{i} \eta-5}{N^4(\mathrm{i}+\eta)^6} \nonumber \\
\begin{split}
&- \frac{2(152 \mathrm{i} \eta^3-99 \eta^2 -42 \mathrm{i} \eta + 7)}{N^5(\mathrm{i}+\eta)^8} 
+\mathcal{O} \left(\frac{1}{N^8} \right).
\end{split}
\label{eq: B parallel correlator semiclassic}
\end{align}
This semiclassical approach of the correlators is valid for $N\gg1$. The alternating sign of the $\eta$-dependent fractions for $D^\times$ and $C^=$ leads to a slow convergence of the polynomials with increasing order $N^{-1}$ for low number of open modes $N$. Instead, $B^=$ shows a faster convergence for low number of open modes. In principle the low $N$-limit brings the semiclassical approximation to its limitation. The universality of the correlators relative to the chosen channels $(a,b,c,d)$ enables in the tight-binding model to average, aside from disorder configurations and $\bar{E}$, also over distinct channel combinations. This guaranties the agreement of the four-point correlators with the expectation from Eqs.~(\ref{eq: D cross correlator semiclassic})-(\ref{eq: B parallel correlator semiclassic})  for ${N=10}$. In Fig.~\ref{fig:four-point corr for N10 KWANT}, the theoretically expected forms are adapted to the numerical results by fitting the dwell time $\tau_D$ and a global multiplicative factor. For all four-point correlators, the resulting dwell time $\tau_D$ agrees with the classical one $\tau_{cl}$ by more than $84 \%$. When reducing the number of open modes to ${N=4}$, as mentioned above the semiclassical polynomials in Eqs.~(\ref{eq: D cross correlator semiclassic})-(\ref{eq: B parallel correlator semiclassic}) oscillate about the limiting curve for large $N$. The formulas adapted to the simulation are shown in Fig.~\ref{fig:four-point corr for N4 KWANT} for different truncations of the $1/N$ sums. Here, $D^\times$ and $B^=$, both having alternating signs, show drastic oscillations [Fig.~\ref{fig:four-point corr for N4 KWANT} (a)] in comparison with $B^=$ in Fig.~\ref{fig:four-point corr for N4 KWANT} (c) and (d). For the simulations enormous $\bar{E}$-averages ensure reliable universal correlators.  

\begin{figure}[ht]
\centering
 \begin{overpic}[width=0.49    \linewidth]{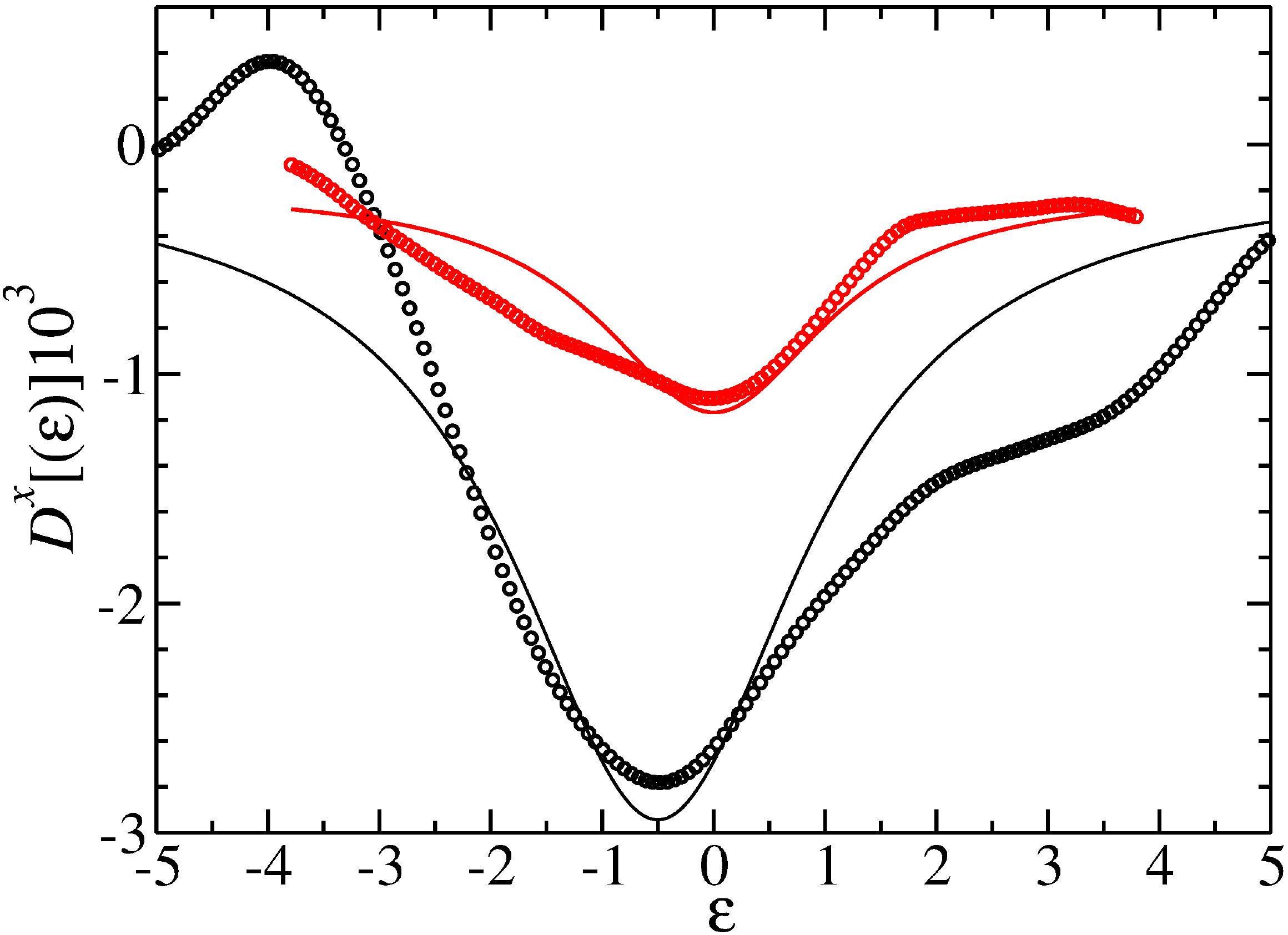}
 \end{overpic}
 \centering
 \begin{overpic}[width=0.49   \linewidth]{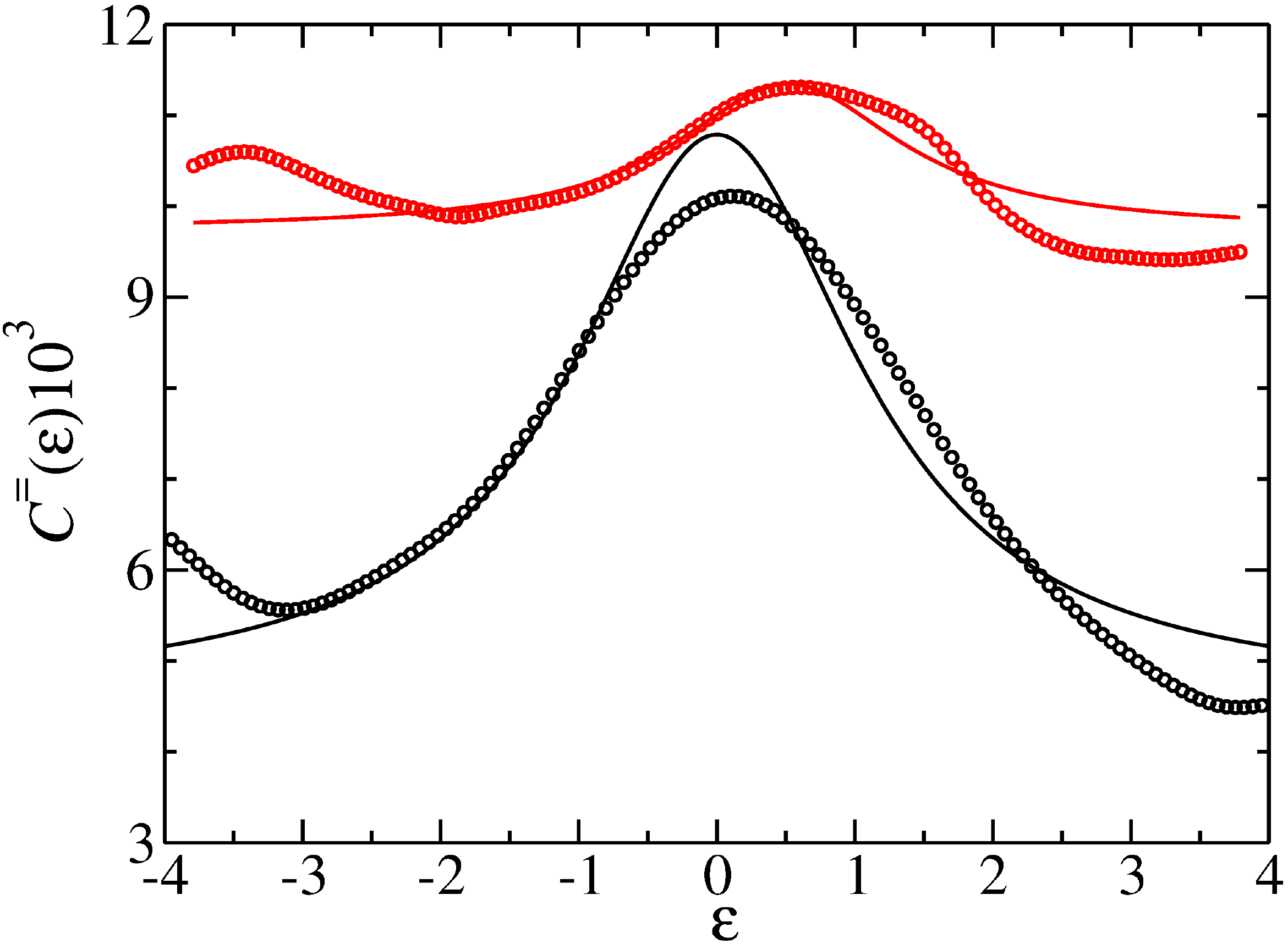}
 \end{overpic}\\
 \centering
 \begin{overpic}[width=0.49    \linewidth]{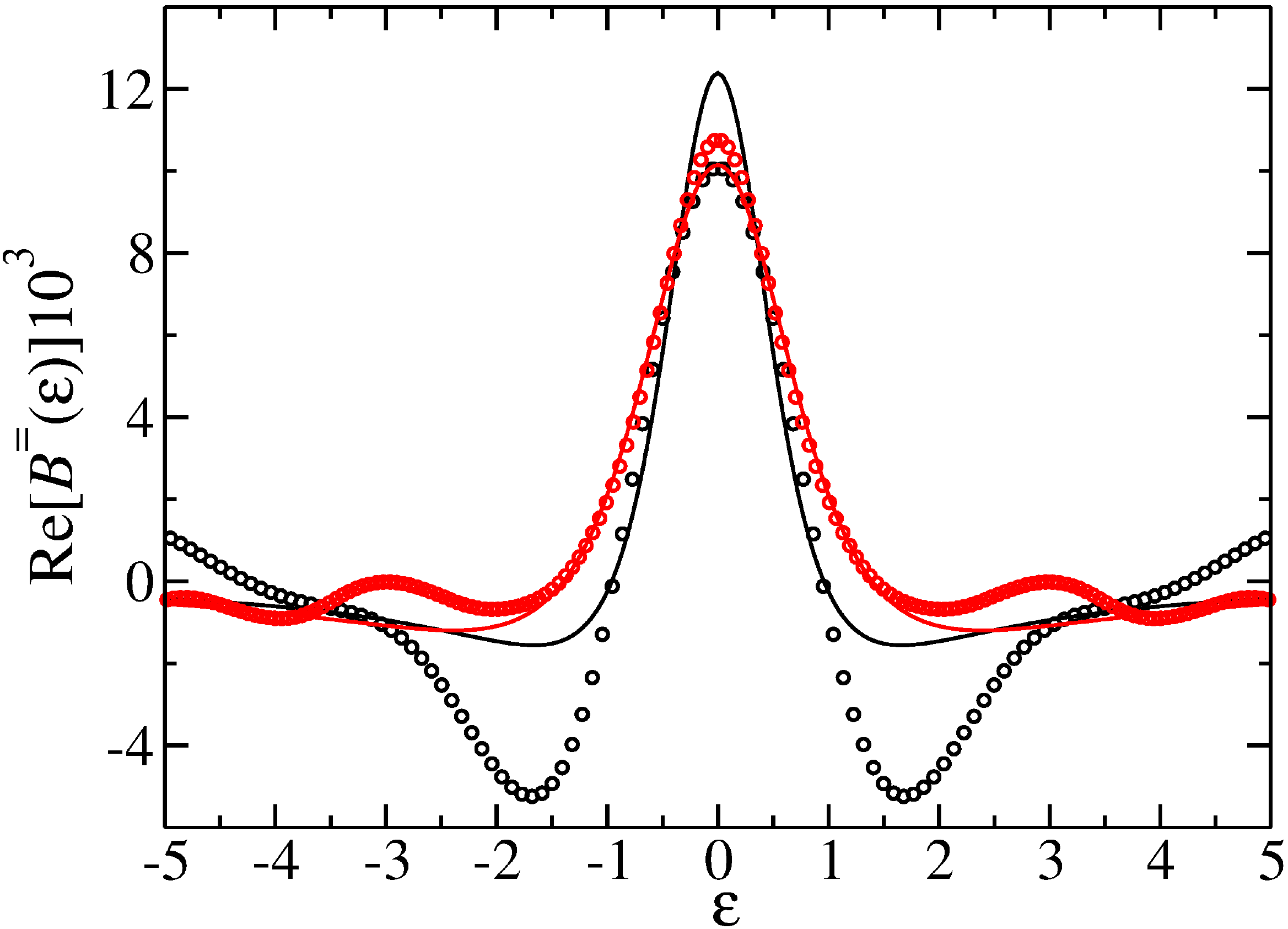}
 \end{overpic}
 \centering
 \begin{overpic}[width=0.49   
 \linewidth]{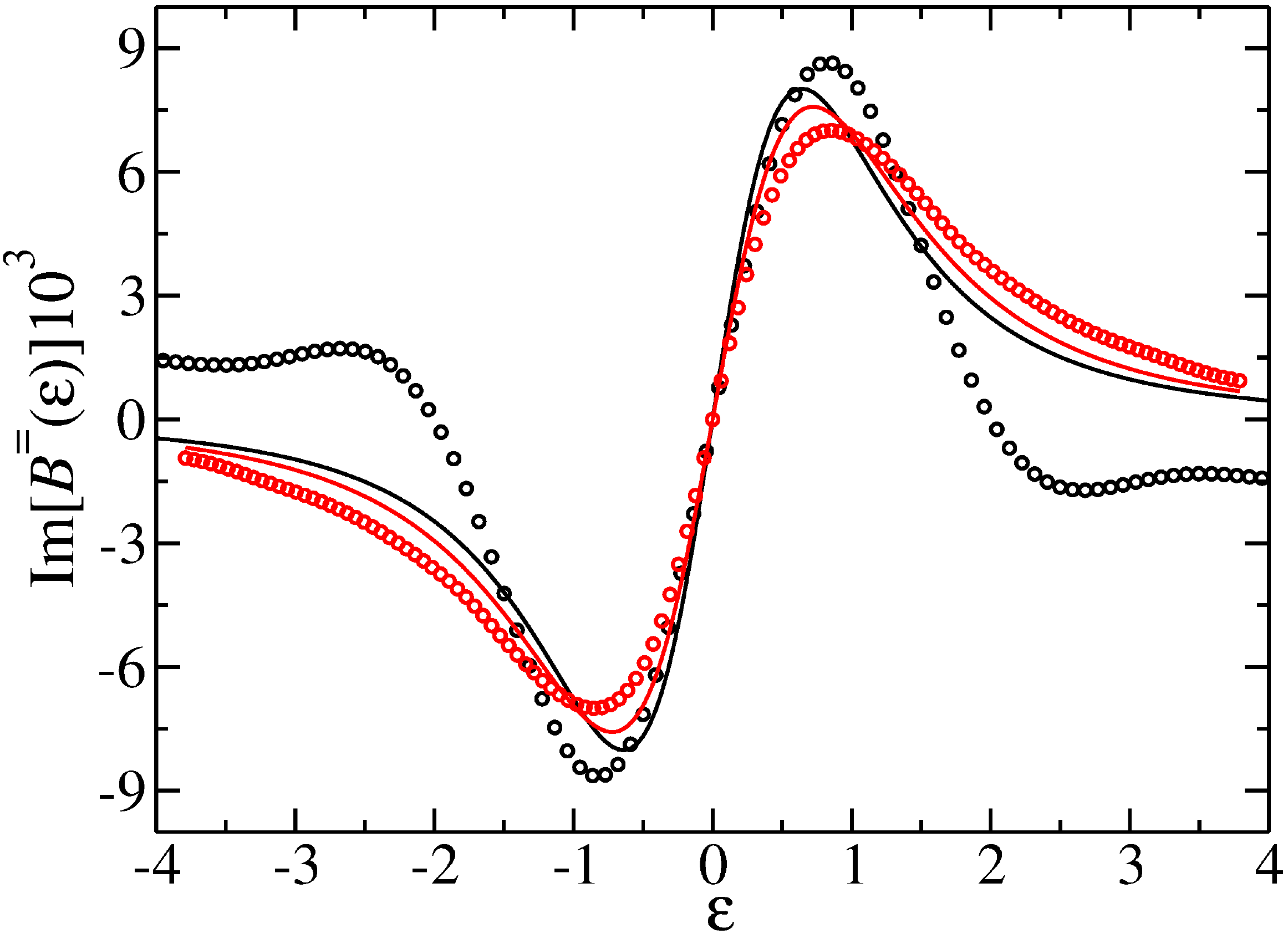}
 \end{overpic}
 
	\caption{The experimental four-point correlators $D^x$, $C^=$, and $B^=$ (black dots) and the best fitting curves deduced from semiclassics, Eqs.~(\ref{eq: D cross correlator semiclassic})-(\ref{eq: B parallel correlator semiclassic}) (black lines). Comparison with RMT simulations for N=9 open channels (red dots) and the corresponding fits to semiclassics (red lines).}
        \label{fig:four-point corr Experiment}
\end{figure}

Compared to that, for the experimental results in Fig.~\ref{fig:four-point corr Experiment} less averaging is necessary: They were obtained by averaging over 30 correlation functions obtained from 6 measurements for each antenna combination in a frequency range of about $0.5$~GHz in the frequency interval ${f\in[10,11.6]}$~GHz to ensure a sufficient number of resonances. Since there is absorption which is unavoidable in the experiments, the curves only agree qualitatively with the predictions. In order to quantify its effect on the S-matrix correlations, we analyzed the properties of the corresponding quantum graph without and with absorption, which was introduced by adding a small imaginary part to the wavenumber $k$~\cite{Kottos1999}; see Sect.~\ref{subsec:Quantum graph with absorption}. These numerical simulations confirmed our assumption that deviations from the analytical results Eqs.~(\ref{eq: D cross correlator semiclassic})-(\ref{eq: B parallel correlator semiclassic}) indeed may be attributed to its presence. We furthermore performed RMT simulations using the Heidelberg approach described in Sect.~\ref{subsec:The Heidelberg approach within RMT}~\cite{Mahaux1966,Dietz2010}. Here we added up to 6 scattering channels to model absorption~\cite{Dietz2010}, i.e., we considered $N=4-10$ open channels, where the transmission coefficients were set equal to unity. We found best agreement between the experimental and RMT correlation functions for $N=9-10$. We also simulated absorption by $\Lambda =50$ fictitious channels~\cite{Dietz2010} with equal transmission coefficients $T_f$ and obtained best results for $\frac{2\pi}{\tilde D}\Gamma_{abs}=\Lambda T_f\simeq 5$, according to the Weisskopf formula~\cite{Blatt1952}. Furthermore, $\Gamma_{abs}$ denotes the contribution of absorption to the resonance widths. 

We found that, except for a possible vertical shift of the correlation function, which generally is eliminated by dividing a correlation function by its value at $\epsilon=0$, the experimental results agree well with the numerical ones for the corresponding quantum graph with absorption. Similar results are obtained for the RMT simulations with $N=4$ open channels plus absorption and with $N=9$ open channels. These studies corroborate that deviations from the predictions may be attributed to absorption. A fit of the analytical results, Eqs.~(\ref{eq: D cross correlator semiclassic})-(\ref{eq: B parallel correlator semiclassic}), with $N$ and $\tau_{QG}$ as fit parameter yielded best agreement for $N=9-10$ open channels and $\tau_{QG}\approx 0.9\tau_D$; see Fig.~\ref{fig:four-point corr Experiment}. For the $D^x$ and the $C^=$ correlators we had to introduce a prefactor as an additional fit parameter in order to obtain a good description of the amplitudes.

\section{Conclusions}
\label{sec:Conclusions}
We have presented a comprehensive study of the S-matrix correlations required in the theoretical description of many-body scattering of non-interacting wavepackets through cavities supporting classically chaotic dynamics. Our study exhausts and compares all known venues for the description of chaotic scattering by addressing the cross energy- and channel- correlations of S-matrix entries using analytical, numerical and experimental approaches. The particular form of the correlators we study is motivated by the microscopic description of the many-body scattering process, and allows us to identify the combinations of single-particle S-matrices that, after suitable averages, are responsible for the emergence of universal features in the many-body transition probabilities. In a first stage, we used two-point correlators as a probe  for the identification of fully chaotic dynamics. This was achieved by comparing the expected universal results of the semiclassical analysis based on interfering paths, Random Matrix Theory simulations, and numerically exact tight-binding results, that are confirmed to exquisite detail by experimental measurements on microwave graphs. 

Once the set of parameters leading to chaotic dynamics is identified, we addressed the universality of the four-point cross-energy correlations required in the study of universal many-body scattering, confirming again the exactness of the semiclassical diagramatic approach by its excellent comparison with numerical simulations and delicate measurements. Of particular importance is our combined study, again using all approaches available, of the convergence properties of the diagramatic semiclassical expansion in the regime of low number of open channels, showing that the asymptotic methods can be indeed pushed into such a regime, a conclusion of importance when addressing the universality aspects of many-body scattering in the framework of electron quantum optics of much present interest.


\textbf{Acknowledgements}.  A.B., J.D.U. and K.R. acknowledge support from Deutsche Forschungsgemeinschaft (DFG,German Research Foundation) within Project-ID 314695032—SFB 1277 (project A07). B.D. thanks the National Natural Science Foundation of China for financial support through Grants Nos. 11775100 and 11961131009. J.C. performed the experiments.


\bibliography{Bib}{}        

\begin{thebibliography}{62}%
\makeatletter
\providecommand \@ifxundefined [1]{%
 \@ifx{#1\undefined}
}%
\providecommand \@ifnum [1]{%
 \ifnum #1\expandafter \@firstoftwo
 \else \expandafter \@secondoftwo
 \fi
}%
\providecommand \@ifx [1]{%
 \ifx #1\expandafter \@firstoftwo
 \else \expandafter \@secondoftwo
 \fi
}%
\providecommand \natexlab [1]{#1}%
\providecommand \enquote  [1]{``#1''}%
\providecommand \bibnamefont  [1]{#1}%
\providecommand \bibfnamefont [1]{#1}%
\providecommand \citenamefont [1]{#1}%
\providecommand \href@noop [0]{\@secondoftwo}%
\providecommand \href [0]{\begingroup \@sanitize@url \@href}%
\providecommand \@href[1]{\@@startlink{#1}\@@href}%
\providecommand \@@href[1]{\endgroup#1\@@endlink}%
\providecommand \@sanitize@url [0]{\catcode `\\12\catcode `\$12\catcode
  `\&12\catcode `\#12\catcode `\^12\catcode `\_12\catcode `\%12\relax}%
\providecommand \@@startlink[1]{}%
\providecommand \@@endlink[0]{}%
\providecommand \url  [0]{\begingroup\@sanitize@url \@url }%
\providecommand \@url [1]{\endgroup\@href {#1}{\urlprefix }}%
\providecommand \urlprefix  [0]{URL }%
\providecommand \Eprint [0]{\href }%
\providecommand \doibase [0]{https://doi.org/}%
\providecommand \selectlanguage [0]{\@gobble}%
\providecommand \bibinfo  [0]{\@secondoftwo}%
\providecommand \bibfield  [0]{\@secondoftwo}%
\providecommand \translation [1]{[#1]}%
\providecommand \BibitemOpen [0]{}%
\providecommand \bibitemStop [0]{}%
\providecommand \bibitemNoStop [0]{.\EOS\space}%
\providecommand \EOS [0]{\spacefactor3000\relax}%
\providecommand \BibitemShut  [1]{\csname bibitem#1\endcsname}%
\let\auto@bib@innerbib\@empty
\bibitem [{\citenamefont {Datta}(1995)}]{datta_1995}%
  \BibitemOpen
  \bibfield  {author} {\bibinfo {author} {\bibfnamefont {S.}~\bibnamefont
  {Datta}},\ }\href {https://doi.org/10.1017/CBO9780511805776} {\emph {\bibinfo
  {title} {Electronic Transport in Mesoscopic Systems}}},\ Cambridge Studies in
  Semiconductor Physics and Microelectronic Engineering\ (\bibinfo  {publisher}
  {Cambridge University Press},\ \bibinfo {year} {1995})\BibitemShut {NoStop}%
\bibitem [{\citenamefont {Imry}(2002)}]{imry2002introduction}%
  \BibitemOpen
  \bibfield  {author} {\bibinfo {author} {\bibfnamefont {Y.}~\bibnamefont
  {Imry}},\ }\href@noop {} {\emph {\bibinfo {title} {Introduction to mesoscopic
  physics}}}\ (\bibinfo  {publisher} {Oxford University Press},\ \bibinfo
  {year} {2002})\BibitemShut {NoStop}%
\bibitem [{\citenamefont {Berry}(1979)}]{Berry1979}%
  \BibitemOpen
  \bibfield  {author} {\bibinfo {author} {\bibfnamefont {M.~V.}\ \bibnamefont
  {Berry}},\ }\href@noop {} {\emph {\bibinfo {title} {Structural Stability in
  Physics}}},\ edited by\ \bibinfo {editor} {\bibfnamefont {W.}~\bibnamefont
  {G{\"u}ttinger}}\ and\ \bibinfo {editor} {\bibfnamefont {H.}~\bibnamefont
  {Eikemeier}}\ (\bibinfo  {publisher} {Springer Berlin Heidelberg},\ \bibinfo
  {address} {Berlin, Heidelberg},\ \bibinfo {year} {1979})\ pp.\ \bibinfo
  {pages} {51--53}\BibitemShut {NoStop}%
\bibitem [{\citenamefont {Casati}\ \emph {et~al.}(1980)\citenamefont {Casati},
  \citenamefont {Valz-Gris},\ and\ \citenamefont {Guarnieri}}]{Casati1980}%
  \BibitemOpen
  \bibfield  {author} {\bibinfo {author} {\bibfnamefont {G.}~\bibnamefont
  {Casati}}, \bibinfo {author} {\bibfnamefont {F.}~\bibnamefont {Valz-Gris}},\
  and\ \bibinfo {author} {\bibfnamefont {I.}~\bibnamefont {Guarnieri}},\
  }\href@noop {} {\bibfield  {journal} {\bibinfo  {journal} {Lettere al Nuovo
  Cimento}\ }\textbf {\bibinfo {volume} {28}},\ \bibinfo {pages} {279}
  (\bibinfo {year} {1980})}\BibitemShut {NoStop}%
\bibitem [{\citenamefont {Bohigas}\ \emph {et~al.}(1984)\citenamefont
  {Bohigas}, \citenamefont {Giannoni},\ and\ \citenamefont
  {Schmit}}]{Bohigas1984}%
  \BibitemOpen
  \bibfield  {author} {\bibinfo {author} {\bibfnamefont {O.}~\bibnamefont
  {Bohigas}}, \bibinfo {author} {\bibfnamefont {M.-J.}\ \bibnamefont
  {Giannoni}},\ and\ \bibinfo {author} {\bibfnamefont {C.}~\bibnamefont
  {Schmit}},\ }\href {https://doi.org/10.1103/PhysRevLett.52.1} {\bibfield
  {journal} {\bibinfo  {journal} {Phys. Rev. Lett.}\ }\textbf {\bibinfo
  {volume} {52}},\ \bibinfo {pages} {1} (\bibinfo {year} {1984})}\BibitemShut
  {NoStop}%
\bibitem [{\citenamefont {Beenakker}(1997)}]{Beenakker1997}%
  \BibitemOpen
  \bibfield  {author} {\bibinfo {author} {\bibfnamefont {C.~W.~J.}\
  \bibnamefont {Beenakker}},\ }\href
  {https://doi.org/10.1103/RevModPhys.69.731} {\bibfield  {journal} {\bibinfo
  {journal} {Rev. Mod. Phys.}\ }\textbf {\bibinfo {volume} {69}},\ \bibinfo
  {pages} {731} (\bibinfo {year} {1997})}\BibitemShut {NoStop}%
\bibitem [{\citenamefont {Richter}(1999)}]{Richter1999}%
  \BibitemOpen
  \bibfield  {author} {\bibinfo {author} {\bibfnamefont {K.}~\bibnamefont
  {Richter}},\ }\href {https://doi.org/10.1007/BFb0109634} {\emph {\bibinfo
  {title} {{Semiclassical Theory of Mesoscopic Quantum Systems}}}}\ (\bibinfo
  {publisher} {Springer, Berlin},\ \bibinfo {year} {1999})\BibitemShut
  {NoStop}%
\bibitem [{\citenamefont {Casati}\ \emph {et~al.}(2000)\citenamefont {Casati},
  \citenamefont {Guarneri},\ and\ \citenamefont {Smilansky}}]{casati2000new}%
  \BibitemOpen
  \bibfield  {author} {\bibinfo {author} {\bibfnamefont {G.}~\bibnamefont
  {Casati}}, \bibinfo {author} {\bibfnamefont {I.}~\bibnamefont {Guarneri}},\
  and\ \bibinfo {author} {\bibfnamefont {U.}~\bibnamefont {Smilansky}},\
  }\href@noop {} {\emph {\bibinfo {title} {New directions in quantum chaos}}},\
  Vol.\ \bibinfo {volume} {143}\ (\bibinfo  {publisher} {IOS Press},\ \bibinfo
  {year} {2000})\BibitemShut {NoStop}%
\bibitem [{\citenamefont {Mahaux}\ and\ \citenamefont
  {Weidenmüller}(1966)}]{Mahaux1966}%
  \BibitemOpen
  \bibfield  {author} {\bibinfo {author} {\bibfnamefont {C.}~\bibnamefont
  {Mahaux}}\ and\ \bibinfo {author} {\bibfnamefont {H.}~\bibnamefont
  {Weidenmüller}},\ }\href
  {https://doi.org/https://doi.org/10.1016/0031-9163(66)90143-0} {\bibfield
  {journal} {\bibinfo  {journal} {Physics Letters}\ }\textbf {\bibinfo {volume}
  {23}},\ \bibinfo {pages} {100 } (\bibinfo {year} {1966})}\BibitemShut
  {NoStop}%
\bibitem [{\citenamefont {Mehta}(2004)}]{mehta2004random}%
  \BibitemOpen
  \bibfield  {author} {\bibinfo {author} {\bibfnamefont {M.~L.}\ \bibnamefont
  {Mehta}},\ }\href@noop {} {\emph {\bibinfo {title} {Random matrices}}}\
  (\bibinfo  {publisher} {Elsevier},\ \bibinfo {year} {2004})\BibitemShut
  {NoStop}%
\bibitem [{\citenamefont {Guhr}\ \emph {et~al.}(1998)\citenamefont {Guhr},
  \citenamefont {M{\"u}ller-Groelling},\ and\ \citenamefont
  {Weidenm{\"u}ller}}]{guhr1998random}%
  \BibitemOpen
  \bibfield  {author} {\bibinfo {author} {\bibfnamefont {T.}~\bibnamefont
  {Guhr}}, \bibinfo {author} {\bibfnamefont {A.}~\bibnamefont
  {M{\"u}ller-Groelling}},\ and\ \bibinfo {author} {\bibfnamefont
  {H.}~\bibnamefont {Weidenm{\"u}ller}},\ }\href@noop {} {\bibfield  {journal}
  {\bibinfo  {journal} {Phys. Rep}\ }\textbf {\bibinfo {volume} {299}},\
  \bibinfo {pages} {190} (\bibinfo {year} {1998})}\BibitemShut {NoStop}%
\bibitem [{\citenamefont {Jalabert}\ \emph {et~al.}(1990)\citenamefont
  {Jalabert}, \citenamefont {Baranger},\ and\ \citenamefont
  {Stone}}]{jalabert1990}%
  \BibitemOpen
  \bibfield  {author} {\bibinfo {author} {\bibfnamefont {R.~A.}\ \bibnamefont
  {Jalabert}}, \bibinfo {author} {\bibfnamefont {H.~U.}\ \bibnamefont
  {Baranger}},\ and\ \bibinfo {author} {\bibfnamefont {A.~D.}\ \bibnamefont
  {Stone}},\ }\href@noop {} {\bibfield  {journal} {\bibinfo  {journal} {Phys.
  Rev. Lett.}\ }\textbf {\bibinfo {volume} {65}},\ \bibinfo {pages} {2442}
  (\bibinfo {year} {1990})}\BibitemShut {NoStop}%
\bibitem [{\citenamefont {Waltner}(2012)}]{waltner2012semiclassical}%
  \BibitemOpen
  \bibfield  {author} {\bibinfo {author} {\bibfnamefont {D.}~\bibnamefont
  {Waltner}},\ }\href@noop {} {\emph {\bibinfo {title} {Semiclassical approach
  to mesoscopic systems: classical trajectory correlations and wave
  interference}}},\ Vol.\ \bibinfo {volume} {245}\ (\bibinfo  {publisher}
  {Springer, Berlin},\ \bibinfo {year} {2012})\BibitemShut {NoStop}%
\bibitem [{\citenamefont {Richter}\ and\ \citenamefont
  {Sieber}(2002)}]{Richter2002}%
  \BibitemOpen
  \bibfield  {author} {\bibinfo {author} {\bibfnamefont {K.}~\bibnamefont
  {Richter}}\ and\ \bibinfo {author} {\bibfnamefont {M.}~\bibnamefont
  {Sieber}},\ }\href {https://doi.org/10.1103/PhysRevLett.89.206801} {\bibfield
   {journal} {\bibinfo  {journal} {Phys. Rev. Lett.}\ }\textbf {\bibinfo
  {volume} {89}},\ \bibinfo {pages} {206801} (\bibinfo {year}
  {2002})}\BibitemShut {NoStop}%
\bibitem [{\citenamefont {Tillmann}\ \emph {et~al.}(2013)\citenamefont
  {Tillmann}, \citenamefont {Daki{\'c}}, \citenamefont {Heilmann},
  \citenamefont {Nolte}, \citenamefont {Szameit},\ and\ \citenamefont
  {Walther}}]{tillmann2013experimental}%
  \BibitemOpen
  \bibfield  {author} {\bibinfo {author} {\bibfnamefont {M.}~\bibnamefont
  {Tillmann}}, \bibinfo {author} {\bibfnamefont {B.}~\bibnamefont {Daki{\'c}}},
  \bibinfo {author} {\bibfnamefont {R.}~\bibnamefont {Heilmann}}, \bibinfo
  {author} {\bibfnamefont {S.}~\bibnamefont {Nolte}}, \bibinfo {author}
  {\bibfnamefont {A.}~\bibnamefont {Szameit}},\ and\ \bibinfo {author}
  {\bibfnamefont {P.}~\bibnamefont {Walther}},\ }\href@noop {} {\bibfield
  {journal} {\bibinfo  {journal} {Nature photonics}\ }\textbf {\bibinfo
  {volume} {7}},\ \bibinfo {pages} {540} (\bibinfo {year} {2013})}\BibitemShut
  {NoStop}%
\bibitem [{\citenamefont {Wang}\ \emph {et~al.}(2017)\citenamefont {Wang},
  \citenamefont {He}, \citenamefont {Li}, \citenamefont {Su}, \citenamefont
  {Li}, \citenamefont {Huang}, \citenamefont {Ding}, \citenamefont {Chen},
  \citenamefont {Liu},\ and\ \citenamefont {Qin}}]{wang2017high}%
  \BibitemOpen
  \bibfield  {author} {\bibinfo {author} {\bibfnamefont {H.}~\bibnamefont
  {Wang}}, \bibinfo {author} {\bibfnamefont {Y.}~\bibnamefont {He}}, \bibinfo
  {author} {\bibfnamefont {Y.-H.}\ \bibnamefont {Li}}, \bibinfo {author}
  {\bibfnamefont {Z.-E.}\ \bibnamefont {Su}}, \bibinfo {author} {\bibfnamefont
  {B.}~\bibnamefont {Li}}, \bibinfo {author} {\bibfnamefont {H.-L.}\
  \bibnamefont {Huang}}, \bibinfo {author} {\bibfnamefont {X.}~\bibnamefont
  {Ding}}, \bibinfo {author} {\bibfnamefont {M.-C.}\ \bibnamefont {Chen}},
  \bibinfo {author} {\bibfnamefont {C.}~\bibnamefont {Liu}},\ and\ \bibinfo
  {author} {\bibfnamefont {J.}~\bibnamefont {Qin}},\ }\href@noop {} {\bibfield
  {journal} {\bibinfo  {journal} {Nature Photonics}\ }\textbf {\bibinfo
  {volume} {11}},\ \bibinfo {pages} {361} (\bibinfo {year} {2017})}\BibitemShut
  {NoStop}%
\bibitem [{\citenamefont {Aaronson}\ and\ \citenamefont
  {Arkhipov}(2011)}]{aaronson2011computational}%
  \BibitemOpen
  \bibfield  {author} {\bibinfo {author} {\bibfnamefont {S.}~\bibnamefont
  {Aaronson}}\ and\ \bibinfo {author} {\bibfnamefont {A.}~\bibnamefont
  {Arkhipov}},\ }in\ \href@noop {} {\emph {\bibinfo {booktitle} {Proceedings of
  the forty-third annual ACM symposium on Theory of computing}}}\ (\bibinfo
  {year} {2011})\ pp.\ \bibinfo {pages} {333--342}\BibitemShut {NoStop}%
\bibitem [{\citenamefont {Urbina}\ \emph {et~al.}(2016)\citenamefont {Urbina},
  \citenamefont {Kuipers}, \citenamefont {Matsumoto}, \citenamefont {Hummel},\
  and\ \citenamefont {Richter}}]{Urbina_2016}%
  \BibitemOpen
  \bibfield  {author} {\bibinfo {author} {\bibfnamefont {J.-D.}\ \bibnamefont
  {Urbina}}, \bibinfo {author} {\bibfnamefont {J.}~\bibnamefont {Kuipers}},
  \bibinfo {author} {\bibfnamefont {S.}~\bibnamefont {Matsumoto}}, \bibinfo
  {author} {\bibfnamefont {Q.}~\bibnamefont {Hummel}},\ and\ \bibinfo {author}
  {\bibfnamefont {K.}~\bibnamefont {Richter}},\ }\href
  {https://doi.org/10.1103/PhysRevLett.116.100401} {\bibfield  {journal}
  {\bibinfo  {journal} {Phys. Rev. Lett.}\ }\textbf {\bibinfo {volume} {116}},\
  \bibinfo {pages} {100401} (\bibinfo {year} {2016})}\BibitemShut {NoStop}%
\bibitem [{\citenamefont {Oliveira}\ and\ \citenamefont
  {Novaes}(2020)}]{oliveira2020immanants}%
  \BibitemOpen
  \bibfield  {author} {\bibinfo {author} {\bibfnamefont {L.~H.}\ \bibnamefont
  {Oliveira}}\ and\ \bibinfo {author} {\bibfnamefont {M.}~\bibnamefont
  {Novaes}},\ }\href@noop {} {\bibfield  {journal} {\bibinfo  {journal} {arXiv
  preprint arXiv:2002.08112}\ } (\bibinfo {year} {2020})}\BibitemShut {NoStop}%
\bibitem [{\citenamefont {Weinberg}(1995)}]{weinberg1995quantum}%
  \BibitemOpen
  \bibfield  {author} {\bibinfo {author} {\bibfnamefont {S.}~\bibnamefont
  {Weinberg}},\ }\href@noop {} {\emph {\bibinfo {title} {The quantum theory of
  fields}}},\ Vol.~\bibinfo {volume} {2}\ (\bibinfo  {publisher} {Cambridge
  university press},\ \bibinfo {year} {1995})\BibitemShut {NoStop}%
\bibitem [{\citenamefont {Newton}(2013)}]{newton2013scattering}%
  \BibitemOpen
  \bibfield  {author} {\bibinfo {author} {\bibfnamefont {R.~G.}\ \bibnamefont
  {Newton}},\ }\href@noop {} {\emph {\bibinfo {title} {Scattering theory of
  waves and particles}}}\ (\bibinfo  {publisher} {Springer Science \& Business
  Media},\ \bibinfo {year} {2013})\BibitemShut {NoStop}%
\bibitem [{\citenamefont {Sakurai}\ and\ \citenamefont
  {Commins}(1995)}]{sakurai1995}%
  \BibitemOpen
  \bibfield  {author} {\bibinfo {author} {\bibfnamefont {J.~J.}\ \bibnamefont
  {Sakurai}}\ and\ \bibinfo {author} {\bibfnamefont {E.~D.}\ \bibnamefont
  {Commins}},\ }\href {https://doi.org/10.1119/1.17781} {\bibfield  {journal}
  {\bibinfo  {journal} {American Journal of Physics}\ }\textbf {\bibinfo
  {volume} {63}},\ \bibinfo {pages} {93} (\bibinfo {year} {1995})}\BibitemShut
  {NoStop}%
\bibitem [{\citenamefont {Tichy}\ \emph {et~al.}(2014)\citenamefont {Tichy},
  \citenamefont {Mayer}, \citenamefont {Buchleitner},\ and\ \citenamefont
  {M{\o}lmer}}]{tichy2014stringent}%
  \BibitemOpen
  \bibfield  {author} {\bibinfo {author} {\bibfnamefont {M.~C.}\ \bibnamefont
  {Tichy}}, \bibinfo {author} {\bibfnamefont {K.}~\bibnamefont {Mayer}},
  \bibinfo {author} {\bibfnamefont {A.}~\bibnamefont {Buchleitner}},\ and\
  \bibinfo {author} {\bibfnamefont {K.}~\bibnamefont {M{\o}lmer}},\ }\href@noop
  {} {\bibfield  {journal} {\bibinfo  {journal} {Phys. Rev. Lett.}\ }\textbf
  {\bibinfo {volume} {113}},\ \bibinfo {pages} {020502} (\bibinfo {year}
  {2014})}\BibitemShut {NoStop}%
\bibitem [{\citenamefont {Ra}\ \emph {et~al.}(2013)\citenamefont {Ra},
  \citenamefont {Tichy}, \citenamefont {Lim}, \citenamefont {Kwon},
  \citenamefont {Mintert}, \citenamefont {Buchleitner},\ and\ \citenamefont
  {Kim}}]{Ra1227}%
  \BibitemOpen
  \bibfield  {author} {\bibinfo {author} {\bibfnamefont {Y.-S.}\ \bibnamefont
  {Ra}}, \bibinfo {author} {\bibfnamefont {M.~C.}\ \bibnamefont {Tichy}},
  \bibinfo {author} {\bibfnamefont {H.-T.}\ \bibnamefont {Lim}}, \bibinfo
  {author} {\bibfnamefont {O.}~\bibnamefont {Kwon}}, \bibinfo {author}
  {\bibfnamefont {F.}~\bibnamefont {Mintert}}, \bibinfo {author} {\bibfnamefont
  {A.}~\bibnamefont {Buchleitner}},\ and\ \bibinfo {author} {\bibfnamefont
  {Y.-H.}\ \bibnamefont {Kim}},\ }\href
  {https://doi.org/10.1073/pnas.1206910110} {\bibfield  {journal} {\bibinfo
  {journal} {Proceedings of the National Academy of Sciences}\ }\textbf
  {\bibinfo {volume} {110}},\ \bibinfo {pages} {1227} (\bibinfo {year}
  {2013})}\BibitemShut {NoStop}%
\bibitem [{\citenamefont {Broome}\ \emph {et~al.}(2013)\citenamefont {Broome},
  \citenamefont {Fedrizzi}, \citenamefont {Rahimi-Keshari}, \citenamefont
  {Dove}, \citenamefont {Aaronson}, \citenamefont {Ralph},\ and\ \citenamefont
  {White}}]{Broome794}%
  \BibitemOpen
  \bibfield  {author} {\bibinfo {author} {\bibfnamefont {M.~A.}\ \bibnamefont
  {Broome}}, \bibinfo {author} {\bibfnamefont {A.}~\bibnamefont {Fedrizzi}},
  \bibinfo {author} {\bibfnamefont {S.}~\bibnamefont {Rahimi-Keshari}},
  \bibinfo {author} {\bibfnamefont {J.}~\bibnamefont {Dove}}, \bibinfo {author}
  {\bibfnamefont {S.}~\bibnamefont {Aaronson}}, \bibinfo {author}
  {\bibfnamefont {T.~C.}\ \bibnamefont {Ralph}},\ and\ \bibinfo {author}
  {\bibfnamefont {A.~G.}\ \bibnamefont {White}},\ }\href
  {https://doi.org/10.1126/science.1231440} {\bibfield  {journal} {\bibinfo
  {journal} {Science}\ }\textbf {\bibinfo {volume} {339}},\ \bibinfo {pages}
  {794} (\bibinfo {year} {2013})}\BibitemShut {NoStop}%
\bibitem [{\citenamefont {Hong}\ \emph {et~al.}(1987)\citenamefont {Hong},
  \citenamefont {Ou},\ and\ \citenamefont {Mandel}}]{Hong1987}%
  \BibitemOpen
  \bibfield  {author} {\bibinfo {author} {\bibfnamefont {C.~K.}\ \bibnamefont
  {Hong}}, \bibinfo {author} {\bibfnamefont {Z.~Y.}\ \bibnamefont {Ou}},\ and\
  \bibinfo {author} {\bibfnamefont {L.}~\bibnamefont {Mandel}},\ }\href
  {https://doi.org/10.1103/PhysRevLett.59.2044} {\bibfield  {journal} {\bibinfo
   {journal} {Phys. Rev. Lett.}\ }\textbf {\bibinfo {volume} {59}},\ \bibinfo
  {pages} {2044} (\bibinfo {year} {1987})}\BibitemShut {NoStop}%
\bibitem [{\citenamefont {Davis}\ and\ \citenamefont
  {Boos{\'e}}(1989)}]{Davis1989}%
  \BibitemOpen
  \bibfield  {author} {\bibinfo {author} {\bibfnamefont {E.}~\bibnamefont
  {Davis}}\ and\ \bibinfo {author} {\bibfnamefont {D.}~\bibnamefont
  {Boos{\'e}}},\ }\href@noop {} {\bibfield  {journal} {\bibinfo  {journal} {Z.
  Phys. A}\ }\textbf {\bibinfo {volume} {332}},\ \bibinfo {pages} {427}
  (\bibinfo {year} {1989})}\BibitemShut {NoStop}%
\bibitem [{\citenamefont {Dietz}\ \emph
  {et~al.}(2010{\natexlab{a}})\citenamefont {Dietz}, \citenamefont {Harney},
  \citenamefont {Richter}, \citenamefont {Sch{\"a}fer},\ and\ \citenamefont
  {Weidenm{\"u}ller}}]{dietz2010cross}%
  \BibitemOpen
  \bibfield  {author} {\bibinfo {author} {\bibfnamefont {B.}~\bibnamefont
  {Dietz}}, \bibinfo {author} {\bibfnamefont {H.}~\bibnamefont {Harney}},
  \bibinfo {author} {\bibfnamefont {A.}~\bibnamefont {Richter}}, \bibinfo
  {author} {\bibfnamefont {F.}~\bibnamefont {Sch{\"a}fer}},\ and\ \bibinfo
  {author} {\bibfnamefont {H.}~\bibnamefont {Weidenm{\"u}ller}},\ }\href@noop
  {} {\bibfield  {journal} {\bibinfo  {journal} {Physics Letters B}\ }\textbf
  {\bibinfo {volume} {685}},\ \bibinfo {pages} {263} (\bibinfo {year}
  {2010}{\natexlab{a}})}\BibitemShut {NoStop}%
\bibitem [{\citenamefont {Landauer}(1970)}]{landauer1970}%
  \BibitemOpen
  \bibfield  {author} {\bibinfo {author} {\bibfnamefont {R.}~\bibnamefont
  {Landauer}},\ }\href@noop {} {\bibfield  {journal} {\bibinfo  {journal}
  {Philosophical magazine}\ }\textbf {\bibinfo {volume} {21}},\ \bibinfo
  {pages} {863} (\bibinfo {year} {1970})}\BibitemShut {NoStop}%
\bibitem [{\citenamefont {B{\"u}ttiker}(1986)}]{buttiker1986}%
  \BibitemOpen
  \bibfield  {author} {\bibinfo {author} {\bibfnamefont {M.}~\bibnamefont
  {B{\"u}ttiker}},\ }\href@noop {} {\bibfield  {journal} {\bibinfo  {journal}
  {Phys. Rev. Lett.}\ }\textbf {\bibinfo {volume} {57}},\ \bibinfo {pages}
  {1761} (\bibinfo {year} {1986})}\BibitemShut {NoStop}%
\bibitem [{\citenamefont {Baranger}\ \emph {et~al.}(1993)\citenamefont
  {Baranger}, \citenamefont {Jalabert},\ and\ \citenamefont
  {Stone}}]{baranger1993}%
  \BibitemOpen
  \bibfield  {author} {\bibinfo {author} {\bibfnamefont {H.~U.}\ \bibnamefont
  {Baranger}}, \bibinfo {author} {\bibfnamefont {R.~A.}\ \bibnamefont
  {Jalabert}},\ and\ \bibinfo {author} {\bibfnamefont {A.~D.}\ \bibnamefont
  {Stone}},\ }\href@noop {} {\bibfield  {journal} {\bibinfo  {journal} {Chaos}\
  }\textbf {\bibinfo {volume} {3}},\ \bibinfo {pages} {665} (\bibinfo {year}
  {1993})}\BibitemShut {NoStop}%
\bibitem [{\citenamefont {Gutzwiller}(2013)}]{gutzwiller2013}%
  \BibitemOpen
  \bibfield  {author} {\bibinfo {author} {\bibfnamefont {M.~C.}\ \bibnamefont
  {Gutzwiller}},\ }\href@noop {} {\emph {\bibinfo {title} {Chaos in classical
  and quantum mechanics}}},\ Vol.~\bibinfo {volume} {1}\ (\bibinfo  {publisher}
  {Springer, Berlin},\ \bibinfo {year} {2013})\BibitemShut {NoStop}%
\bibitem [{\citenamefont {Fisher}\ and\ \citenamefont
  {Lee}(1981)}]{Fisher1981}%
  \BibitemOpen
  \bibfield  {author} {\bibinfo {author} {\bibfnamefont {D.~S.}\ \bibnamefont
  {Fisher}}\ and\ \bibinfo {author} {\bibfnamefont {P.~A.}\ \bibnamefont
  {Lee}},\ }\href {https://doi.org/10.1103/PhysRevB.23.6851} {\bibfield
  {journal} {\bibinfo  {journal} {Phys. Rev. B}\ }\textbf {\bibinfo {volume}
  {23}},\ \bibinfo {pages} {6851} (\bibinfo {year} {1981})}\BibitemShut
  {NoStop}%
\bibitem [{\citenamefont {Baranger}\ and\ \citenamefont
  {Mello}(1996)}]{Baranger1996}%
  \BibitemOpen
  \bibfield  {author} {\bibinfo {author} {\bibfnamefont {H.~U.}\ \bibnamefont
  {Baranger}}\ and\ \bibinfo {author} {\bibfnamefont {P.~A.}\ \bibnamefont
  {Mello}},\ }\href {https://doi.org/10.1103/PhysRevB.54.R14297} {\bibfield
  {journal} {\bibinfo  {journal} {Phys. Rev. B}\ }\textbf {\bibinfo {volume}
  {54}},\ \bibinfo {pages} {R14297} (\bibinfo {year} {1996})}\BibitemShut
  {NoStop}%
\bibitem [{\citenamefont {Kuipers}(2009)}]{Kuipers_2009}%
  \BibitemOpen
  \bibfield  {author} {\bibinfo {author} {\bibfnamefont {J.}~\bibnamefont
  {Kuipers}},\ }\href {https://doi.org/10.1088/1751-8113/42/42/425101}
  {\bibfield  {journal} {\bibinfo  {journal} {Journal of Physics A:
  Mathematical and Theoretical}\ }\textbf {\bibinfo {volume} {42}},\ \bibinfo
  {pages} {425101} (\bibinfo {year} {2009})}\BibitemShut {NoStop}%
\bibitem [{\citenamefont {Novaes}(2016)}]{Novaes_2corr}%
  \BibitemOpen
  \bibfield  {author} {\bibinfo {author} {\bibfnamefont {M.}~\bibnamefont
  {Novaes}},\ }\href {https://doi.org/10.1063/1.4972288} {\bibfield  {journal}
  {\bibinfo  {journal} {Journal of Mathematical Physics}\ }\textbf {\bibinfo
  {volume} {57}},\ \bibinfo {pages} {122105} (\bibinfo {year}
  {2016})}\BibitemShut {NoStop}%
\bibitem [{\citenamefont {Ericson}\ \emph {et~al.}(2016)\citenamefont
  {Ericson}, \citenamefont {Dietz},\ and\ \citenamefont
  {Richter}}]{Ericson2016}%
  \BibitemOpen
  \bibfield  {author} {\bibinfo {author} {\bibfnamefont {T.~E.}\ \bibnamefont
  {Ericson}}, \bibinfo {author} {\bibfnamefont {B.}~\bibnamefont {Dietz}},\
  and\ \bibinfo {author} {\bibfnamefont {A.}~\bibnamefont {Richter}},\ }\href
  {https://doi.org/10.1103/PhysRevE.94.042207} {\bibfield  {journal} {\bibinfo
  {journal} {Phys. Rev. E}\ }\textbf {\bibinfo {volume} {94}},\ \bibinfo
  {pages} {1} (\bibinfo {year} {2016})}\BibitemShut {NoStop}%
\bibitem [{\citenamefont {M\"uller}\ \emph {et~al.}(2007)\citenamefont
  {M\"uller}, \citenamefont {Heusler}, \citenamefont {Braun},\ and\
  \citenamefont {Haake}}]{M_ller_2007}%
  \BibitemOpen
  \bibfield  {author} {\bibinfo {author} {\bibfnamefont {S.}~\bibnamefont
  {M\"uller}}, \bibinfo {author} {\bibfnamefont {S.}~\bibnamefont {Heusler}},
  \bibinfo {author} {\bibfnamefont {P.}~\bibnamefont {Braun}},\ and\ \bibinfo
  {author} {\bibfnamefont {F.}~\bibnamefont {Haake}},\ }\href
  {https://doi.org/10.1088/1367-2630/9/1/012} {\bibfield  {journal} {\bibinfo
  {journal} {New Journal of Physics}\ }\textbf {\bibinfo {volume} {9}},\
  \bibinfo {pages} {12} (\bibinfo {year} {2007})}\BibitemShut {NoStop}%
\bibitem [{\citenamefont {Berkolaiko}\ and\ \citenamefont
  {Kuipers}(2012)}]{Berkolaiko2012}%
  \BibitemOpen
  \bibfield  {author} {\bibinfo {author} {\bibfnamefont {G.}~\bibnamefont
  {Berkolaiko}}\ and\ \bibinfo {author} {\bibfnamefont {J.}~\bibnamefont
  {Kuipers}},\ }\href {https://doi.org/10.1103/PhysRevE.85.045201} {\bibfield
  {journal} {\bibinfo  {journal} {Phys. Rev. E}\ }\textbf {\bibinfo {volume}
  {85}},\ \bibinfo {pages} {045201} (\bibinfo {year} {2012})}\BibitemShut
  {NoStop}%
\bibitem [{\citenamefont {Berkolaiko}\ and\ \citenamefont
  {Kuipers}(2013)}]{Berkolaiko2013}%
  \BibitemOpen
  \bibfield  {author} {\bibinfo {author} {\bibfnamefont {G.}~\bibnamefont
  {Berkolaiko}}\ and\ \bibinfo {author} {\bibfnamefont {J.}~\bibnamefont
  {Kuipers}},\ }\href {https://doi.org/10.1063/1.4842375} {\bibfield  {journal}
  {\bibinfo  {journal} {Journal of Mathematical Physics}\ }\textbf {\bibinfo
  {volume} {54}},\ \bibinfo {pages} {123505} (\bibinfo {year}
  {2013})}\BibitemShut {NoStop}%
\bibitem [{\citenamefont {Kottos}\ and\ \citenamefont
  {Smilansky}(1997)}]{Kottos1997}%
  \BibitemOpen
  \bibfield  {author} {\bibinfo {author} {\bibfnamefont {T.}~\bibnamefont
  {Kottos}}\ and\ \bibinfo {author} {\bibfnamefont {U.}~\bibnamefont
  {Smilansky}},\ }\href {https://doi.org/10.1103/PhysRevLett.79.4794}
  {\bibfield  {journal} {\bibinfo  {journal} {Phys. Rev. Lett.}\ }\textbf
  {\bibinfo {volume} {79}},\ \bibinfo {pages} {4794} (\bibinfo {year}
  {1997})}\BibitemShut {NoStop}%
\bibitem [{\citenamefont {Kottos}\ and\ \citenamefont
  {Smilansky}(1999)}]{Kottos1999}%
  \BibitemOpen
  \bibfield  {author} {\bibinfo {author} {\bibfnamefont {T.}~\bibnamefont
  {Kottos}}\ and\ \bibinfo {author} {\bibfnamefont {U.}~\bibnamefont
  {Smilansky}},\ }\href
  {https://doi.org/https://doi.org/10.1006/aphy.1999.5904} {\bibfield
  {journal} {\bibinfo  {journal} {Annals of Physics}\ }\textbf {\bibinfo
  {volume} {274}},\ \bibinfo {pages} {76} (\bibinfo {year} {1999})}\BibitemShut
  {NoStop}%
\bibitem [{\citenamefont {Pakonski}\ \emph {et~al.}(2001)\citenamefont
  {Pakonski}, \citenamefont {Zyczkowski},\ and\ \citenamefont
  {Kus}}]{Pakonski2001}%
  \BibitemOpen
  \bibfield  {author} {\bibinfo {author} {\bibfnamefont {P.}~\bibnamefont
  {Pakonski}}, \bibinfo {author} {\bibfnamefont {K.}~\bibnamefont
  {Zyczkowski}},\ and\ \bibinfo {author} {\bibfnamefont {M.}~\bibnamefont
  {Kus}},\ }\href {https://doi.org/10.1088/0305-4470/34/43/313} {\bibfield
  {journal} {\bibinfo  {journal} {Journal of Physics A: Mathematical and
  General}\ }\textbf {\bibinfo {volume} {34}},\ \bibinfo {pages} {9303}
  (\bibinfo {year} {2001})}\BibitemShut {NoStop}%
\bibitem [{\citenamefont {Texier}\ and\ \citenamefont
  {Montambaux}(2001)}]{Texier2001}%
  \BibitemOpen
  \bibfield  {author} {\bibinfo {author} {\bibfnamefont {C.}~\bibnamefont
  {Texier}}\ and\ \bibinfo {author} {\bibfnamefont {G.}~\bibnamefont
  {Montambaux}},\ }\href {https://doi.org/10.1088/0305-4470/34/47/328}
  {\bibfield  {journal} {\bibinfo  {journal} {J. Phys. A}\ }\textbf {\bibinfo
  {volume} {34}},\ \bibinfo {pages} {10307} (\bibinfo {year}
  {2001})}\BibitemShut {NoStop}%
\bibitem [{\citenamefont {St\"{o}ckmann}(1999)}]{Stoeckmann1999}%
  \BibitemOpen
  \bibfield  {author} {\bibinfo {author} {\bibfnamefont {H.-J.}\ \bibnamefont
  {St\"{o}ckmann}},\ }\href {https://doi.org/10.1017/CBO9780511524622} {\emph
  {\bibinfo {title} {Quantum chaos, an introduction}}}\ (\bibinfo  {publisher}
  {Cambridge University Press, Cambridge},\ \bibinfo {year} {1999})\BibitemShut
  {NoStop}%
\bibitem [{\citenamefont {Haake}(2001)}]{Haake2001}%
  \BibitemOpen
  \bibfield  {author} {\bibinfo {author} {\bibfnamefont {F.}~\bibnamefont
  {Haake}},\ }\href {https://doi.org/10.1007/978-3-662-04506-0} {\emph
  {\bibinfo {title} {Quantum signatures of chaos}}}\ (\bibinfo  {publisher}
  {Springer-Verlag, Berlin},\ \bibinfo {year} {2001})\BibitemShut {NoStop}%
\bibitem [{\citenamefont {Gnutzmann}\ and\ \citenamefont
  {Altland}(2004)}]{Gnutzmann2004}%
  \BibitemOpen
  \bibfield  {author} {\bibinfo {author} {\bibfnamefont {S.}~\bibnamefont
  {Gnutzmann}}\ and\ \bibinfo {author} {\bibfnamefont {A.}~\bibnamefont
  {Altland}},\ }\href {https://doi.org/10.1103/PhysRevLett.93.194101}
  {\bibfield  {journal} {\bibinfo  {journal} {Phys. Rev. Lett.}\ }\textbf
  {\bibinfo {volume} {93}},\ \bibinfo {pages} {194101} (\bibinfo {year}
  {2004})}\BibitemShut {NoStop}%
\bibitem [{\citenamefont {Mehta}(1991)}]{Mehta1990}%
  \BibitemOpen
  \bibfield  {author} {\bibinfo {author} {\bibfnamefont {M.~L.}\ \bibnamefont
  {Mehta}},\ }\href@noop {} {\emph {\bibinfo {title} {Random matrices}}}\
  (\bibinfo  {publisher} {Academic Press, Inc., Boston, MA},\ \bibinfo {year}
  {1991})\BibitemShut {NoStop}%
\bibitem [{\citenamefont {Pluha{\v{r}}}\ and\ \citenamefont
  {Weidenm{\"u}ller}(2013{\natexlab{a}})}]{Pluhar2013}%
  \BibitemOpen
  \bibfield  {author} {\bibinfo {author} {\bibfnamefont {Z.}~\bibnamefont
  {Pluha{\v{r}}}}\ and\ \bibinfo {author} {\bibfnamefont {H.~A.}\ \bibnamefont
  {Weidenm{\"u}ller}},\ }\href@noop {} {\bibfield  {journal} {\bibinfo
  {journal} {Phys. Rev. Lett.}\ }\textbf {\bibinfo {volume} {110}},\ \bibinfo
  {pages} {034101} (\bibinfo {year} {2013}{\natexlab{a}})}\BibitemShut
  {NoStop}%
\bibitem [{\citenamefont {Pluha{\v{r}}}\ and\ \citenamefont
  {Weidenm{\"u}ller}(2013{\natexlab{b}})}]{Pluhar2013a}%
  \BibitemOpen
  \bibfield  {author} {\bibinfo {author} {\bibfnamefont {Z.}~\bibnamefont
  {Pluha{\v{r}}}}\ and\ \bibinfo {author} {\bibfnamefont {H.}~\bibnamefont
  {Weidenm{\"u}ller}},\ }\href@noop {} {\bibfield  {journal} {\bibinfo
  {journal} {Phys. Rev. E}\ }\textbf {\bibinfo {volume} {88}},\ \bibinfo
  {pages} {022902} (\bibinfo {year} {2013}{\natexlab{b}})}\BibitemShut
  {NoStop}%
\bibitem [{\citenamefont {Pluha{\v{r}}}\ and\ \citenamefont
  {Weidenm{\"u}ller}(2014)}]{Pluhar2014}%
  \BibitemOpen
  \bibfield  {author} {\bibinfo {author} {\bibfnamefont {Z.}~\bibnamefont
  {Pluha{\v{r}}}}\ and\ \bibinfo {author} {\bibfnamefont {H.}~\bibnamefont
  {Weidenm{\"u}ller}},\ }\href@noop {} {\bibfield  {journal} {\bibinfo
  {journal} {Phys. Rev. Lett.}\ }\textbf {\bibinfo {volume} {112}},\ \bibinfo
  {pages} {144102} (\bibinfo {year} {2014})}\BibitemShut {NoStop}%
\bibitem [{\citenamefont {Hul}\ \emph {et~al.}(2004)\citenamefont {Hul},
  \citenamefont {Bauch}, \citenamefont {Pako\ifmmode~\acute{n}\else
  \'{n}\fi{}ski}, \citenamefont {Savytskyy}, \citenamefont
  {\ifmmode~\dot{Z}\else \.{Z}\fi{}yczkowski},\ and\ \citenamefont
  {Sirko}}]{Hul2004}%
  \BibitemOpen
  \bibfield  {author} {\bibinfo {author} {\bibfnamefont {O.}~\bibnamefont
  {Hul}}, \bibinfo {author} {\bibfnamefont {S.}~\bibnamefont {Bauch}}, \bibinfo
  {author} {\bibfnamefont {P.}~\bibnamefont {Pako\ifmmode~\acute{n}\else
  \'{n}\fi{}ski}}, \bibinfo {author} {\bibfnamefont {N.}~\bibnamefont
  {Savytskyy}}, \bibinfo {author} {\bibfnamefont {K.}~\bibnamefont
  {\ifmmode~\dot{Z}\else \.{Z}\fi{}yczkowski}},\ and\ \bibinfo {author}
  {\bibfnamefont {L.}~\bibnamefont {Sirko}},\ }\href
  {https://doi.org/10.1103/PhysRevE.69.056205} {\bibfield  {journal} {\bibinfo
  {journal} {Phys. Rev. E}\ }\textbf {\bibinfo {volume} {69}},\ \bibinfo
  {pages} {056205} (\bibinfo {year} {2004})}\BibitemShut {NoStop}%
\bibitem [{\citenamefont {{\L}awniczak}\ \emph {et~al.}(2010)\citenamefont
  {{\L}awniczak}, \citenamefont {Bauch}, \citenamefont {Hul},\ and\
  \citenamefont {Sirko}}]{Lawniczak2010}%
  \BibitemOpen
  \bibfield  {author} {\bibinfo {author} {\bibfnamefont {M.}~\bibnamefont
  {{\L}awniczak}}, \bibinfo {author} {\bibfnamefont {S.}~\bibnamefont {Bauch}},
  \bibinfo {author} {\bibfnamefont {O.}~\bibnamefont {Hul}},\ and\ \bibinfo
  {author} {\bibfnamefont {L.}~\bibnamefont {Sirko}},\ }\href@noop {}
  {\bibfield  {journal} {\bibinfo  {journal} {Phys. Rev. E}\ }\textbf {\bibinfo
  {volume} {81}},\ \bibinfo {pages} {046204} (\bibinfo {year}
  {2010})}\BibitemShut {NoStop}%
\bibitem [{\citenamefont {Allgaier}\ \emph {et~al.}(2014)\citenamefont
  {Allgaier}, \citenamefont {Gehler}, \citenamefont {Barkhofen}, \citenamefont
  {St{\"o}ckmann},\ and\ \citenamefont {Kuhl}}]{Allgaier2014}%
  \BibitemOpen
  \bibfield  {author} {\bibinfo {author} {\bibfnamefont {M.}~\bibnamefont
  {Allgaier}}, \bibinfo {author} {\bibfnamefont {S.}~\bibnamefont {Gehler}},
  \bibinfo {author} {\bibfnamefont {S.}~\bibnamefont {Barkhofen}}, \bibinfo
  {author} {\bibfnamefont {H.-J.}\ \bibnamefont {St{\"o}ckmann}},\ and\
  \bibinfo {author} {\bibfnamefont {U.}~\bibnamefont {Kuhl}},\ }\href@noop {}
  {\bibfield  {journal} {\bibinfo  {journal} {Phys. Rev. E}\ }\textbf {\bibinfo
  {volume} {89}},\ \bibinfo {pages} {022925} (\bibinfo {year}
  {2014})}\BibitemShut {NoStop}%
\bibitem [{\citenamefont {Bia{\l}ous}\ \emph {et~al.}(2016)\citenamefont
  {Bia{\l}ous}, \citenamefont {Yunko}, \citenamefont {Bauch}, \citenamefont
  {{\L}awniczak}, \citenamefont {Dietz},\ and\ \citenamefont
  {Sirko}}]{Bialous2016}%
  \BibitemOpen
  \bibfield  {author} {\bibinfo {author} {\bibfnamefont {M.}~\bibnamefont
  {Bia{\l}ous}}, \bibinfo {author} {\bibfnamefont {V.}~\bibnamefont {Yunko}},
  \bibinfo {author} {\bibfnamefont {S.}~\bibnamefont {Bauch}}, \bibinfo
  {author} {\bibfnamefont {M.}~\bibnamefont {{\L}awniczak}}, \bibinfo {author}
  {\bibfnamefont {B.}~\bibnamefont {Dietz}},\ and\ \bibinfo {author}
  {\bibfnamefont {L.}~\bibnamefont {Sirko}},\ }\href@noop {} {\bibfield
  {journal} {\bibinfo  {journal} {Phys. Rev. Lett.}\ }\textbf {\bibinfo
  {volume} {117}},\ \bibinfo {pages} {144101} (\bibinfo {year}
  {2016})}\BibitemShut {NoStop}%
\bibitem [{\citenamefont {Rehemanjiang}\ \emph {et~al.}(2016)\citenamefont
  {Rehemanjiang}, \citenamefont {Allgaier}, \citenamefont {Joyner},
  \citenamefont {M{\"u}ller}, \citenamefont {Sieber}, \citenamefont {Kuhl},\
  and\ \citenamefont {St{\"o}ckmann}}]{Rehemanjiang2016}%
  \BibitemOpen
  \bibfield  {author} {\bibinfo {author} {\bibfnamefont {A.}~\bibnamefont
  {Rehemanjiang}}, \bibinfo {author} {\bibfnamefont {M.}~\bibnamefont
  {Allgaier}}, \bibinfo {author} {\bibfnamefont {C.}~\bibnamefont {Joyner}},
  \bibinfo {author} {\bibfnamefont {S.}~\bibnamefont {M{\"u}ller}}, \bibinfo
  {author} {\bibfnamefont {M.}~\bibnamefont {Sieber}}, \bibinfo {author}
  {\bibfnamefont {U.}~\bibnamefont {Kuhl}},\ and\ \bibinfo {author}
  {\bibfnamefont {H.-J.}\ \bibnamefont {St{\"o}ckmann}},\ }\href@noop {}
  {\bibfield  {journal} {\bibinfo  {journal} {Phys. Rev. Lett.}\ }\textbf
  {\bibinfo {volume} {117}},\ \bibinfo {pages} {064101} (\bibinfo {year}
  {2016})}\BibitemShut {NoStop}%
\bibitem [{\citenamefont {Lu}\ \emph {et~al.}(2020)\citenamefont {Lu},
  \citenamefont {Che}, \citenamefont {Zhang},\ and\ \citenamefont
  {Dietz}}]{Lu2020}%
  \BibitemOpen
  \bibfield  {author} {\bibinfo {author} {\bibfnamefont {J.}~\bibnamefont
  {Lu}}, \bibinfo {author} {\bibfnamefont {J.}~\bibnamefont {Che}}, \bibinfo
  {author} {\bibfnamefont {X.}~\bibnamefont {Zhang}},\ and\ \bibinfo {author}
  {\bibfnamefont {B.}~\bibnamefont {Dietz}},\ }\href
  {https://doi.org/10.1103/PhysRevE.102.022309} {\bibfield  {journal} {\bibinfo
   {journal} {Phys. Rev. E}\ }\textbf {\bibinfo {volume} {102}},\ \bibinfo
  {pages} {022309} (\bibinfo {year} {2020})}\BibitemShut {NoStop}%
\bibitem [{\citenamefont {Agassi}\ \emph {et~al.}(1975)\citenamefont {Agassi},
  \citenamefont {Weidenm\"uller},\ and\ \citenamefont
  {Mantzouranis}}]{Agassi1975}%
  \BibitemOpen
  \bibfield  {author} {\bibinfo {author} {\bibfnamefont {D.}~\bibnamefont
  {Agassi}}, \bibinfo {author} {\bibfnamefont {H.}~\bibnamefont
  {Weidenm\"uller}},\ and\ \bibinfo {author} {\bibfnamefont {G.}~\bibnamefont
  {Mantzouranis}},\ }\href
  {https://doi.org/https://doi.org/10.1016/0370-1573(75)90028-9} {\bibfield
  {journal} {\bibinfo  {journal} {Phys. Rep.}\ }\textbf {\bibinfo {volume}
  {22}},\ \bibinfo {pages} {145} (\bibinfo {year} {1975})}\BibitemShut
  {NoStop}%
\bibitem [{\citenamefont {Kumar}\ \emph {et~al.}(2013)\citenamefont {Kumar},
  \citenamefont {Nock}, \citenamefont {Sommers}, \citenamefont {Guhr},
  \citenamefont {Dietz}, \citenamefont {Miski-Oglu}, \citenamefont {Richter},\
  and\ \citenamefont {Sch\"afer}}]{Kumar2013}%
  \BibitemOpen
  \bibfield  {author} {\bibinfo {author} {\bibfnamefont {S.}~\bibnamefont
  {Kumar}}, \bibinfo {author} {\bibfnamefont {A.}~\bibnamefont {Nock}},
  \bibinfo {author} {\bibfnamefont {H.-J.}\ \bibnamefont {Sommers}}, \bibinfo
  {author} {\bibfnamefont {T.}~\bibnamefont {Guhr}}, \bibinfo {author}
  {\bibfnamefont {B.}~\bibnamefont {Dietz}}, \bibinfo {author} {\bibfnamefont
  {M.}~\bibnamefont {Miski-Oglu}}, \bibinfo {author} {\bibfnamefont
  {A.}~\bibnamefont {Richter}},\ and\ \bibinfo {author} {\bibfnamefont
  {F.}~\bibnamefont {Sch\"afer}},\ }\href
  {https://doi.org/10.1103/PhysRevLett.111.030403} {\bibfield  {journal}
  {\bibinfo  {journal} {Phys. Rev. Lett.}\ }\textbf {\bibinfo {volume} {111}},\
  \bibinfo {pages} {030403} (\bibinfo {year} {2013})}\BibitemShut {NoStop}%
\bibitem [{\citenamefont {Groth}\ \emph {et~al.}(2014)\citenamefont {Groth},
  \citenamefont {Wimmer}, \citenamefont {Akhmerov},\ and\ \citenamefont
  {Waintal}}]{Groth_2014}%
  \BibitemOpen
  \bibfield  {author} {\bibinfo {author} {\bibfnamefont {C.~W.}\ \bibnamefont
  {Groth}}, \bibinfo {author} {\bibfnamefont {M.}~\bibnamefont {Wimmer}},
  \bibinfo {author} {\bibfnamefont {A.~R.}\ \bibnamefont {Akhmerov}},\ and\
  \bibinfo {author} {\bibfnamefont {X.}~\bibnamefont {Waintal}},\ }\href
  {https://doi.org/10.1088/1367-2630/16/6/063065} {\bibfield  {journal}
  {\bibinfo  {journal} {New Journal of Physics}\ }\textbf {\bibinfo {volume}
  {16}},\ \bibinfo {pages} {063065} (\bibinfo {year} {2014})}\BibitemShut
  {NoStop}%
\bibitem [{\citenamefont {Dietz}\ \emph
  {et~al.}(2010{\natexlab{b}})\citenamefont {Dietz}, \citenamefont {Friedrich},
  \citenamefont {Harney}, \citenamefont {Miski-Oglu}, \citenamefont {Richter},
  \citenamefont {Sch\"afer},\ and\ \citenamefont {Weidenm\"uller}}]{Dietz2010}%
  \BibitemOpen
  \bibfield  {author} {\bibinfo {author} {\bibfnamefont {B.}~\bibnamefont
  {Dietz}}, \bibinfo {author} {\bibfnamefont {T.}~\bibnamefont {Friedrich}},
  \bibinfo {author} {\bibfnamefont {H.~L.}\ \bibnamefont {Harney}}, \bibinfo
  {author} {\bibfnamefont {M.}~\bibnamefont {Miski-Oglu}}, \bibinfo {author}
  {\bibfnamefont {A.}~\bibnamefont {Richter}}, \bibinfo {author} {\bibfnamefont
  {F.}~\bibnamefont {Sch\"afer}},\ and\ \bibinfo {author} {\bibfnamefont
  {H.~A.}\ \bibnamefont {Weidenm\"uller}},\ }\href
  {https://doi.org/10.1103/PhysRevE.81.036205} {\bibfield  {journal} {\bibinfo
  {journal} {Phys. Rev. E}\ }\textbf {\bibinfo {volume} {81}},\ \bibinfo
  {pages} {036205} (\bibinfo {year} {2010}{\natexlab{b}})}\BibitemShut
  {NoStop}%
\bibitem [{\citenamefont {Blatt}\ \emph {et~al.}(1953)\citenamefont {Blatt},
  \citenamefont {Weisskopf},\ and\ \citenamefont {Dyson}}]{Blatt1952}%
  \BibitemOpen
  \bibfield  {author} {\bibinfo {author} {\bibfnamefont {J.~M.}\ \bibnamefont
  {Blatt}}, \bibinfo {author} {\bibfnamefont {V.~F.}\ \bibnamefont
  {Weisskopf}},\ and\ \bibinfo {author} {\bibfnamefont {F.~J.}\ \bibnamefont
  {Dyson}},\ }\href {https://doi.org/10.1063/1.3061164} {\bibfield  {journal}
  {\bibinfo  {journal} {Physics Today}\ }\textbf {\bibinfo {volume} {6}},\
  \bibinfo {pages} {17} (\bibinfo {year} {1953})}\BibitemShut {NoStop}%
\end{thebibliography}%

\end{document}